# TECHNICAL, ORGANIZATIONAL AND ORAL HISTORY REGARDING THE SOIL SAMPLES MEASUREMENTS FOR CS-137 BECAUSE OF THE CHERNOBYL ACCIDENT FALLOUT (in Greek)

**N.P. Petropoulos**
**Nuclear Engineering Laboratory, School of Mechanical Engineering,**
**National Technical University of Athens, 15780 Athens, GREECE**
**e-mail: npetr@mail.ntua.gr**

**Abstract**

Data are given, commentary is supplied and explanations are provided with regard to the technical, the organizational and, of course, the human history connected to the time of research, which resulted to the paper entitled *"Soil sampling and Cs-137 analysis of the Chernobyl fallout in Greece"*, written by late Professor S.E. Simopoulos. This paper has been provided in Greek translation within an issued honorary volume (ISBN 978-960-254-714-4). Reasonably, the narration starts with the review of the political, the financial and the social situation of Greece around 1986. Subsequently, an analysis is given on the then available means, the persons involved, the methods used, the lessons learned and any other connection with the oral history of the NTUA's Nuclear Engineering Laboratory and other relevant Greek Laboratories. For this history, written proof is now scarce and the persons available to pass it on are growing less and less. N.P. Petropoulos, now Laboratory member and then student of Professor S.E. Simopoulos was in charge of preparation of this text.

*Keywords: Chernobyl, Greece*

# ΤΕΧΝΙΚΗ, ΟΡΓΑΝΩΤΙΚΗ ΚΑΙ ΠΡΟΦΟΡΙΚΗ ΙΣΤΟΡΙΑ ΣΧΕΤΙΚΑ ΜΕ ΤΙΣ ΜΕΤΡΗΣΕΙΣ ΔΕΙΓΜΑΤΩΝ ΕΔΑΦΟΥΣ ΓΙΑ CS-137 ΑΠΟ ΤΗΝ ΕΝΑΠΟΘΕΣΗ ΕΞΑΙΤΙΑΣ ΤΟΥ ΑΤΥΧΗΜΑΤΟΣ ΣΤΟ CHERNOBYL

**Ν.Π. Πετρόπουλος**
**Εργαστήριο Πυρηνικής Τεχνολογίας, Σχολή Μηχανολόγων Μηχανικών,**
**Εθνικό Μετσόβιο Πολυτεχνείο, 15780 Αθήνα**
**e-mail: npetr@mail.ntua.gr**

**Περίληψη**

Δίνονται στοιχεία, γίνονται σχόλια και παρέχονται επεξηγήσεις που συνδέονται με την τεχνική, την οργανωτική και ασφαλώς και την ανθρώπινη ιστορία που αφορά στην εποχή της έρευνας για την εργασία του εκλιπόντος Καθηγητή Σ.Ε. Σιμόπουλου με θέμα *"Δειγματοληψία εδαφών και ανάλυσή τους για Cs-137 από την εναπόθεση εξαιτίας του ατυχήματος στο Cher-* *nobyl"*, η απόδοση στα ελληνικά του οποίου δίνεται στο πλαίσιο εκδοθέντος τιμητικού τόμου (ISBN 978-960-254-714-4). Η διήγηση ξεκινάει εύλογα από την αναφορά στο πολιτικό, το οικονομικό και το κοινωνικό περίγραμμα περί το 1986. Ακολούθως αναλύονται τα μέσα που διατέθηκαν, τα πρόσωπα που ενεπλάκησαν, οι τεχνικές που χρησιμοποιήθηκαν, τα μαθήματα που προέκυψαν και ό,τι άλλο συνδέεται με τη σχετική προφορική ιστορία του Εργαστηρίου Πυρηνικής Τεχνολογίας του ΕΜΠ, και άλλων σχετικών Ελληνικών Εργαστηρίων, για την οποία οι γραπτές πηγές είναι πλέον λίγες και τα πρόσωπα που τη μεταδίδουν γίνονται λιγότερα. Την επιμέλεια του περιεχομένου ανέλαβε το μέλος του Εργαστηρίου και μαθητής του Καθηγητή Σ.Ε. Σιμόπουλου, Ν.Π. Πετρόπουλος.

*Λέξεις-Κλειδιά: Chernobyl, Ελλάδα*

## 1. Εισαγωγή

Η επιτυχημένη αποτύπωση της ρύπανσης από ένα βιομηχανικό ατύχημα σε οποιαδήποτε χώρα είναι από μόνη της μία σημαντική πρόκληση. Πόσο μάλλον, αν το ατύχημα από το οποίο προέρχεται η ρύπανση συνέβη σε έναν πυρηνικό αντιδραστήρα άλλης χώρας, ευρισκόμενης αρκετές χιλιάδες χιλιόμετρα μακρύτερα. Η πρώτη και πολύ δικαιολογημένα αναμενόμενη αντίδραση, είναι η αδιαφορία λόγω της απόστασης. Όταν, και αν εγκαίρως, ξεπερασθεί αυτό το στάδιο η αντίδραση συνήθως μεταμορφώνεται σε ελλιπή βιαστική έρευνα που υστερεί σε επαγγελματισμό και ικανοποιητική λεπτομέρεια. Για την επίτευξη επαγγελματισμού και έγκυρων αποτελεσμάτων που μπορούν να σταθούν χωρίς αμφισβήτηση, όπως π.χ. αυτά της εργασίας Simopoulos (1989), πρέπει να συνδράμουν πολλοί παράγοντες που συνδέονται με, για να αναφερθούν μερικοί, τα διαθέσιμα τεχνικά μέσα, το υπάρχον εξειδικευμένο ανθρώπινο δυναμικό, την ερευνητική (και ακαδημαϊκή) ανεξαρτησία, την επάρκεια των πόρων, τη βούληση της διοίκησης και της Πολιτείας, τις πολιτικές και οικονομικές οριακές συνθήκες. Στην εδώ συνοπτική ιστορική ανασκόπηση καταβλήθηκε προσπάθεια να εξετασθούν ολιστικά και σε αλληλοσύνδεση, τα χαρακτηριστικά με βάση τα οποία ολοκληρώθηκαν οι σχετικές έρευνες και ειδικότερα αυτή του Simopoulos (1989). Πολλά από τα αναγκαία για το σκοπό αυτό στοιχεία αντλήθηκαν από το αρχειακό υλικό του Εργαστηρίου Πυρηνικής Τεχνολογίας του ΕΜΠ (ΕΠΤ-ΕΜΠ), καθώς και από την προφορική ιστορία που γνωρίζουν και διηγούνται τα σημερινά (2022) μέλη του.

## 2. Η Ελλάδα μεταξύ 1986 και 1989

### 2.1 Η πολιτική κατάσταση

Το διάστημα μεταξύ Απριλίου 1986, μήνας που συνέβη το ατύχημα στο Chernobyl, και τον Ιούλιο του 1989, μήνας που δημοσιεύθηκε η ερευνητική εργασία του Σ.Ε. Σιμόπουλου με θέμα *"Δειγματοληψία εδαφών και ανάλυσή τους για Cs-137 από την εναπόθεση εξαιτίας του ατυ-* *χήματος στο Chernobyl"*, δεν ήταν μία ήρεμη περίοδος για τη χώρα. Μετά τις βουλευτικές εκλογές της 2 Ιουνίου του 1985, την ευθύνη διακυβέρνησης της χώρας είχε το ΠΑΣΟΚ. Η πολιτική κατάσταση ήταν σχετικά φορτισμένη κυρίως λόγω και του θορύβου που είχε προκαλέσει τον Μάρτιο του 1985 η εκλογή του Προέδρου της Δημοκρατίας. Υπενθυμίζονται επιπλέον και τα ακόλουθα κύρια γεγονότα: 2 Απριλίου 1986 *Έκρηξη βόμβας σε πτήση της TWA προς Αθήνα (4 νεκροί, το αεροπλάνο προσγειώθηκε με ασφάλεια)*, 26 Απριλίου 1986: *Ατύχημα στο Chernobyl*, 13 Σεπτεμβρίου 1986: *Καταστροφικός σεισμός στην Καλαμάτα*, Μάρτιος 1987: *Ένταση στις σχέσεις με την Τουρκία με αφορμή το ερευνητικό σκάφος "Σισμίκ"*, Ιούλιος 1988: *Έναρξη της διερεύνησης του σκανδάλου Κοσκωτά*, Σεπτέμβριος 1988: *Σοβαρή εγχείριση ανοιχτής καρδιάς του τότε πρωθυπουργού Α.Γ. Παπανδρέου*, 18 Ιουνίου 1989: *Βουλευτικές εκλογές*. Η ακολουθία των παραπάνω έντονων πολιτικών γεγονότων δεν επέτρεψε στο γεγονός του ατυχήματος στο Chernobyl και των πιθανών συνεπειών του στην Ελλάδα να κυριαρχήσει για πολύ στην επικαιρότητα. Οπωσδήποτε, σε αυτό συνέβαλε και η δομή της πληροφόρησης και των μέσων μαζικής ενημέρωσης (ΜΜΕ) εκείνης της εποχής. Υπενθυμίζεται ότι υπήρχαν μόνο δύο δίαυλοι δημόσιας τηλεόρασης, ένα δίκτυο δημόσιων ραδιοφωνικών σταθμών κρατικής εμβέλειας, περιφερειακοί ραδιοφωνικοί σταθμοί ιδιοκτησίας των αυτοδιοικητικών αρχών και φυσικά ένα πλήθος από εφημερίδες και άλλα έντυπα με σχετικά υψηλή κυκλοφορία. Εννοείται ότι το διαδίκτυο ήταν κάτι άγνωστο.

### 2.2 Η οικονομική κατάσταση

Τα κύρια χαρακτηριστικά της οικονομίας την υπόψη περίοδο μπορούν να συνοψισθούν ως εξής: (α) Αύξηση του δημόσιου χρέους: Χονδρικά το χρέος αυξήθηκε από περίπου 50% του ΑΕΠ το 1985 σε περίπου 80% του ΑΕΠ το 1986. Το δημόσιο χρέος χρηματοδοτήθηκε κυρίως από εσωτερικό και δευτερευόντως από εξωτερικό δανεισμό. (β) Συνεχίσθηκε αλλά σε πολύ μικρότερη έκταση και με φθίνοντα τρόπο ένα πρόγραμμα κρατικοποιήσεων μεγάλων επιχειρήσεων που είχε ξεκινήσει το διάστημα 1981 - 1985. (γ) Συνεχίσθηκε η σταδιακή αποκλιμάκωση του πληθωρισμού από περίπου 23% σε περίπου 13%. Πρέπει να σημειωθεί ότι χονδρικά και χωρίς να ληφθεί υπόψη ο πληθωρισμός, την περίοδο αυτή το ΑΕΠ της χώρας ήταν τρεις φορές χαμηλότερο από το σημερινό. Επομένως μπορεί να γίνει κατανοητό ότι τα σχετικά μέσα και εργαλεία της χώρας στους περισσότερους τομείς ήταν τουλάχιστον τρεις φορές λιγότερα από ό,τι σήμερα (2022).

### 2.3 Η άσκηση της διοίκησης

Ο χειρισμός των συνεπειών ενός πυρηνικού ατυχήματος όπως αυτό στο Chernobyl, όπως και κάθε πυρηνικού ατυχήματος, είναι ένα πολύπλοκο θέμα για την οποιαδήποτε διοίκηση. Για την Ελληνική Διοίκηση, η οποία σε κυβερνητικό επίπεδο είναι γενικά περισσότερο κατακερματισμένη από ό,τι σε άλλες χώρες, το θέμα του χειρισμού ενός τέτοιου ατυχήματος δυσκολεύει διότι αναπτύσσονται συναρμοδιότητες που πολλές φορές εμποδίζουν τη λειτουργία τόσο της κοινής λογικής, όσο και των σχετικών εργαλείων διαχείρισης της ανησυχίας του κοινού. Αναλύοντας την κυβερνητική δομή εκείνης της περιόδου 1986 - 1989 διαπιστώνονται συναρμοδιότητες στα Υπουργεία με αντικείμενα: Εσωτερικών, Γεωργίας, Περιβάλλοντος, Έρευνας, Υγείας και Δημόσιας Τάξεως. Επιπλέον την υπόψη περίοδο τα πρόσωπα που άσκησαν τη διοίκηση σε επίπεδο υπουργού στα παραπάνω αντικείμενα άλλαξαν στα περισσότερα από αυτά δύο φορές. Οπωσδήποτε η πίεση για τη διερεύνηση των συνεπειών του ατυχήματος τους πρώτους δυο μήνες, τον Μάιο και τον Ιούνιο 1986, υπήρξε τόσο έντονη που έφερε σε αμηχανία τη διοίκηση και ταλαιπώρησε τους όποιους μηχανισμούς υπήρχαν διαθέσιμοι για να παρέχουν τις αναγκαίες απαντήσεις.

### 2.4 Το πολιτικό και το τεχνολογικό διακύβευμα του ατυχήματος

Από πολιτική άποψη πολλές χώρες και ιδιαίτερα αυτές που ήταν γεωγραφικά ή ιδεολογικά ή οικονομικά κοντά στην τότε ΕΣΣΔ, ίσως θα προτιμούσαν να περιορίσουν την έκταση της πληροφόρησης για το ατύχημα στο Chernobyl και να υποβαθμίσουν τις συνέπειές του. Αντίθετα εκείνες που δεν είχαν τέτοιες γειτνιάσεις, ίσως θα προτιμούσαν να υπερτονίσουν το ατύχημα για να αναδείξουν, ίσως και με κάποια υστεροβουλία, εσφαλμένες πολιτικές, τεχνικές και οικονομικές επιλογές. Το κλίμα αυτό αποτυπώθηκε και στην Ελλάδα κυρίως διά μέσω του τύπου, όπως φαίνεται και στις παρακάτω Εικόνες 1 έως και 3.

Υπάρχουν ασφαλώς και τα δύο παραδείγματα. Είναι γνωστό ότι αρκετές (αλλά όχι όλες) από τις χώρες που ανήκαν στο Σύμφωνο της Βαρσοβίας δεν ασχολήθηκαν σε κανένα επίπεδο για τη διερεύνηση της εναπόθεσης ισοτόπων από το ατύχημα στο έδαφός τους ή την είσοδό τους στις τροφές παραγωγής τους για να μην διαταράξουν τις σχέσεις τους με την τότε υπερδύναμη. Η Ελλάδα, κυρίως χάρη στην επιστημονική εργασία των ερευνητικών της κέντρων και των Πανεπιστημιακών της Εργαστηρίων (μεταξύ αυτών ασφαλώς και του Εργαστηρίου Πυρηνικής Τεχνολογίας του ΕΜΠ) και την επιμονή τους στα διαπιστωμένα δεδομένα από προσεκτικές μετρήσεις, κατόρθωσε να έχει ως Πολιτεία σύντομη και ορθή πληροφόρηση σχετικά με τη ρύπανση που προκλήθηκε, την οποία κοινοποίησε στους αρμόδιους Διεθνείς Οργανισμούς χωρίς να χρειασθεί να πάρει πολιτική θέση. Σε αυτό βοήθησε και το γεγονός ότι οι συνέπειες του ατυχήματος στη χώρα υπήρξαν, όπως αποδείχθηκε κάπως αργότερα (Γεροντίδου, 1994), χωρίς σημασία σε επίπεδο υγείας του πληθυσμού ακόμα και για τις περιοχές που τον Μάιο του 1986 βρέθηκαν να παρουσιάζουν την υψηλότερη ραδιενέργεια εδάφους.

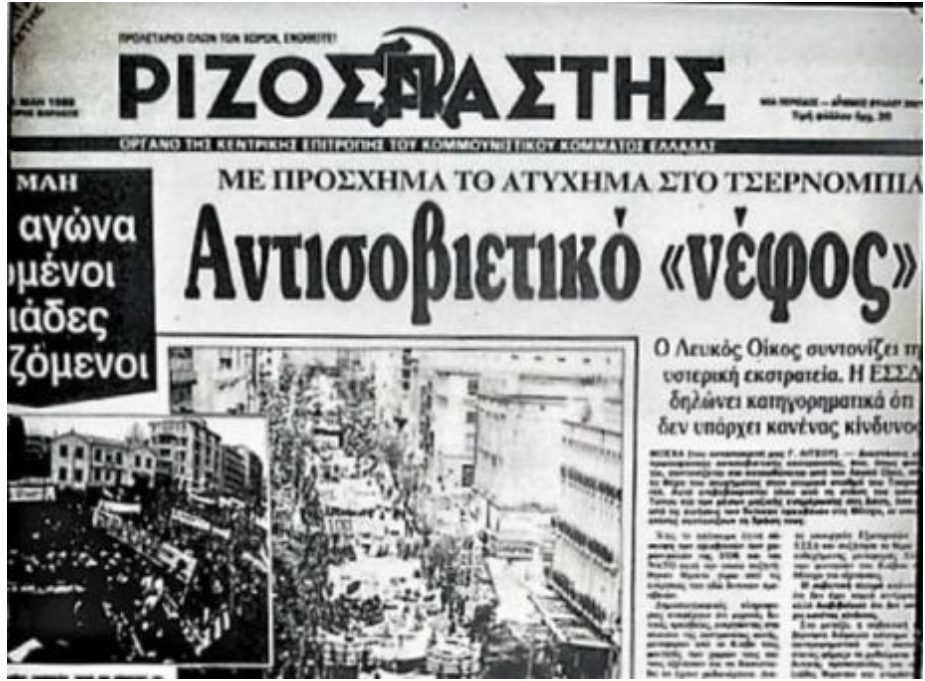

Εικόνα 1. Ένα από τα σχετικά πρωτοσέλιδα του *Ριζοσπάστη* με σαφή φιλοσοβιετική στάση (από το διαδίκτυο)

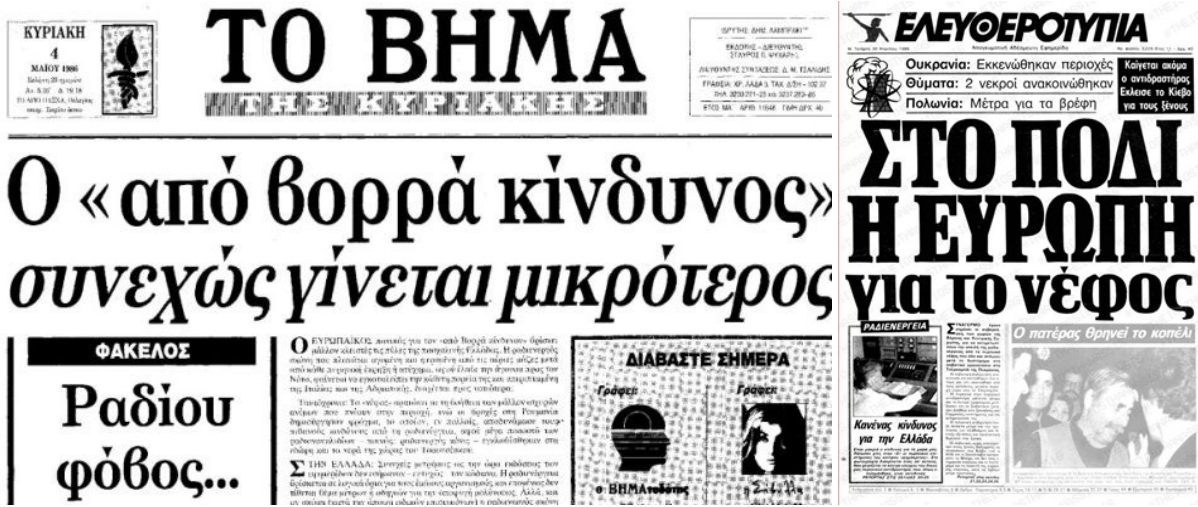

Εικόνα 2. Μάλλον ουδέτερη στάση από τις συμπολιτευόμενες εφημερίδες *Βήμα* και *Ελευθεροτυπία* με σχετικά κεντρώο προσανατολισμό (από το διαδίκτυο)

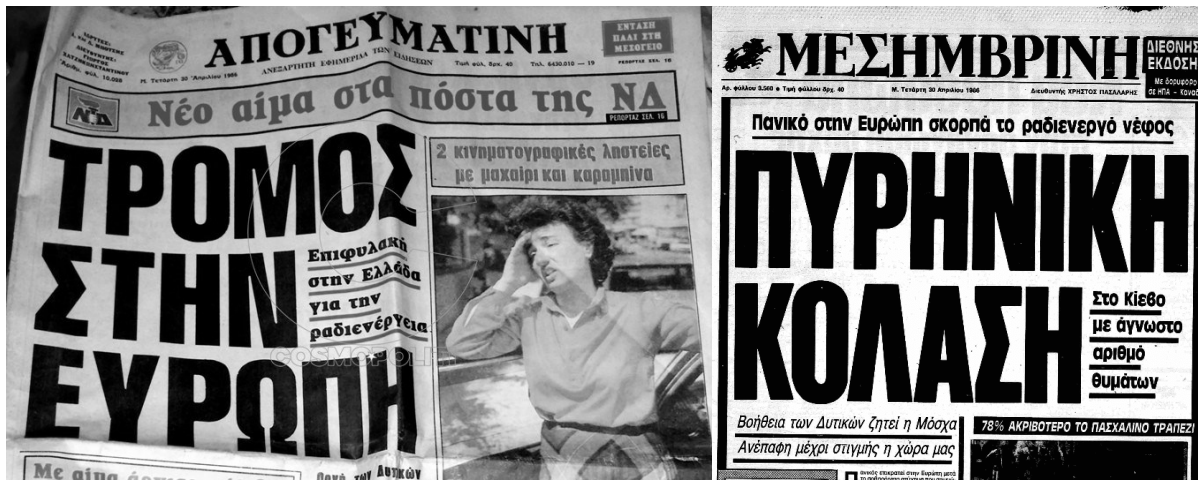

Εικόνα 3. Θορυβώδης στάση από τις αντιπολιτευόμενες εφημερίδες *Απογευματινή* και *Μεσημβρινή* (από το διαδίκτυο)

Τα ΜΜΕ σε εκείνες τις χώρες που βασίζονταν στην πυρηνική τεχνολογία για την παραγωγή ηλεκτρικής ενέργειας ή είχαν έντονα θετική πολιτική και τεχνολογική στάση απέναντι στην πυρηνική τεχνολογία, απέφυγαν να σχολιάσουν σε μεγάλη έκταση και με αρνητικό τρόπο το ατύχημα στο Chernobyl ακόμα και αν στο χώρο τους διαπιστώθηκαν συνέπειες από το ατύχημα. Αυτό έγινε για να αποφευχθεί η ανησυχία του πληθυσμού και η κατάρρευση της εμπιστοσύνης στην υπόψη τεχνολογία. Χαρακτηριστική περίπτωση τέτοιας χώρας είναι η Γαλλία. Από την άλλη, σε χώρες που δεν βασίζονταν στην ηλεκτροπαραγωγή από πυρηνική ενέργεια, ακόμα και αν είχαν θετική πολιτική και τεχνολογική στάση, τα ΜΜΕ δεν δίστασαν να προάγουν την αρνητική δημοσιότητα τη σχετική με το ατύχημα ακόμα και αν στο χώρο τους δεν διαπιστώθηκαν συνέπειες από το ατύχημα. Χαρακτηριστική περίπτωση τέτοιας χώρας είναι η Ιταλία, η οποία μάλιστα, μετά το ατύχημα στο Chernobyl εγκατέλειψε, με δημοψήφισμα, τον προγραμματισμό της να εγκαινιάσει τον πρώτο της πυρηνικό αντιδραστήρα πριν το τέλος της δεκαετίας του 1980 καθώς και οποιονδήποτε άλλο σχετικό μελλοντικό προγραμματισμό. Τέλος, στις χώρες, οι οποίες ήταν κεντρικά πολιτικά τοποθετημένες αντίθετα από την ΕΣΣΔ, κυρίως δηλαδή στις ΗΠΑ και το Ηνωμένο Βασίλειο, παρόλο που αυτές χρησιμοποιούσαν έντονα την πυρηνική τεχνολογία για την παραγωγή πυρηνικής ενέργειας, τα ΜΜΕ δεν δίστασαν να προβάλλουν το ατύχημα με έντονα αρνητικό τρόπο, σημειώνοντας πάντως δίκαια, ότι αυτό έγινε κυρίως επειδή η τεχνολογία και ο τρόπος λειτουργίας του αντιδραστήρα που εξερράγη ήταν λιγότερο από μέτρια. Στην Ελλάδα, ήδη από τις αρχές της δεκαετίας του '80 είχε εξαντληθεί μια συζήτηση σχετικά με την υιοθέτηση της πυρηνικής τεχνολογίας, η οποία είχε ξεκινήσει στα μέσα της δεκαετίας του '60, φθάνοντας σε αρνητικό πρόσημο. Οι κυριότερες αιτίες συνδέονταν με το κόστος του εγχειρήματος για μια χώρα με περιορισμένα οικονομικά μέσα όπως η Ελλάδα, αλλά και με τους προβληματισμούς που προκάλεσε το ατύχημα στο Three Mile Island των ΗΠΑ το 1979. Ως εκ τούτου, από τεχνολογική άποψη η χώρα δεν ζημιώθηκε από την αρνητική δημοσιότητα που έφερε το ατύχημα για την πυρηνική τεχνολογία. Συγκυριακά μάλιστα η χώρα μας στάθηκε τυχερή, διότι αν είχε ξεκινήσει να προετοιμάζεται για έναν πυρηνικό αντιδραστήρα, πιθανότατα, μετά το ατύχημα στο Chernobyl και σε συνδυασμό με τις νέες αντιρρήσεις από την κοινή γνώμη, θα εγκατέλειπε το εγχείρημα, χάνοντας και τα χρήματα της αντίστοιχης επένδυσης.

### 3. Το σχετικό ανθρώπινο και τεχνικό δυναμικό της χώρας το 1986

#### 3.1 Τα Ελληνικά Εργαστήρια με δυνατότητες μετρήσεων

Η διασπορά ισοτόπων εξαιτίας του ατυχήματος στο Chernobyl, ξεπέρασε κάθε τι αναμενόμενο από οποιονδήποτε στην Ευρώπη. Η αιτία για αυτό ήταν ασφαλώς η πυρκαγιά στον γραφίτη του πυρήνα του πυρηνικού αντιδραστήρα που καταστράφηκε, η οποία προκάλεσε τη δημιουργία νέφους με ρύπους που προωθήθηκε σε μεγάλο ύψος στην ανώτερη ατμόσφαιρα και στη συνέχεια παρασύρθηκε από τους ανέμους για να καταπέσει σταδιακά πολύ μακριά από τη γειτονιά του Chernobyl. Για την Ελλάδα, ως πλησιέστερος πυρηνικός κίνδυνος θεωρούνταν πάντοτε οι τέσσερεις παλαιάς τεχνολογίας πυρηνικοί αντιδραστήρες της Βουλγαρίας τύπου VVER-440/V-230 στη θέση Kozloduy, οι οποίοι και σταμάτησαν να λειτουργούν λίγο πριν την είσοδο της χώρας αυτής στην Ευρωπαϊκή Ένωση το 2007. Παρόλα αυτά, ακόμα και για αυτούς τους αντιδραστήρες, όταν λειτουργούσαν, κανείς Έλληνας ή άλλης εθνικότητας ειδικός δεν υποστήριζε ότι τυχόν ατύχημα θα είχε συνέπειες πέραν από την ευρύτερη γειτονιά της θέσης Kozloduy εντός της ΒΔ Βουλγαρίας και ίσως σε κοντινές περιοχές της Ουγγαρίας και της Ρουμανίας. Κατά συνέπεια οι μηχανισμοί της Ελλάδας για τη διαπίστωση συνεπειών από πυρηνικό ατύχημα, χωρίς να είναι πολύ χειρότεροι από εκείνους χωρών ανάλογων οικονομικών δυνατοτήτων, δεν ήταν ασφαλώς οι κατάλληλοι για την έκταση που πήρε το ατύχημα στο Chernobyl και στη χώρα μας. Πιο αναλυτικά μπορεί να σημειωθεί ότι τα κυριότερα διαθέσιμα Εργαστήρια με τον κατάλληλο εξοπλισμό ήταν τα παρακάτω: (α) Εργαστήριο Ραδιενέργειας Περιβάλλοντος (ΕΡΠ) στο Εθνικό Ερευνητικό Κέντρο "Δημόκριτος" (κυριότερος υπεύθυνος Δρ. Π. Κρητίδης), (β) Εργαστήριο Πυρηνικής Τεχνολογίας στο Εθνικό Μετσόβιο Πολυτεχνείο (υπεύθυνοι Δρ. Μ.Γ. Αγγελόπουλος, Δρ. Δ.Ι. Λεωνίδου και Δρ. Σ.Ε. Σιμόπουλος), (γ) Εργαστήριο Πυρηνικής Φυσικής στο Αριστοτέλειο Πανεπιστήμιο Θεσσαλονίκης (ΑΠΘ - κυριότερος υπεύθυνος Δρ. Κ. Παπαστεφάνου), (δ) Εργαστήριο Πυρηνικής Φυσικής στο Πανεπιστήμιο Ιωαννίνων (κυριότερος υπεύθυνος Δρ. Π. Ασημακόπουλος) και (ε) Εργαστήριο Πυρηνικής Τεχνολογίας στο Αριστοτέλειο Πανεπιστήμιο Θεσσαλονίκης (υπεύθυνος μετρήσεων Δρ. Α. Κλούβας). Πρέπει να σημειωθεί ότι αυτά τα Εργαστήρια είχαν περίπου ισοδύναμο εξοπλισμό σε διαφορετικό βαθμό ετοιμότητας και λειτουργικότητας και όχι πάντα συνδυασμένο με το κατάλληλο προσωπικό. Για παράδειγμα αναφέρεται ότι, ασφαλώς χάρη και στις προσπάθειες του Σ.Ε. Σιμόπουλου, αλλά και άλλων που θα αναφερθούν, ο εξοπλισμός του Εργαστηρίου Πυρηνικής Τεχνολογίας του ΕΜΠ, την περίοδο μετά το 1980 ήταν πάντοτε σε ετοιμότητα και σε λειτουργική κατάσταση, ως σύνολο. Δεν συμπεριλαμβάνονται στον κατάλογο αυτό μικρότερα Εργαστήρια με λιγότερο εξοπλισμό, όπως π.χ. αυτά του Πανεπιστημίου Αθηνών, ή του Πανεπιστημίου Πατρών ή του Γεωπονικού Πανεπιστημίου. Το βάρος των αρχικών μετρήσεων των σχετικών με το ατύχημα σήκωσε πρακτικά μόνο του το Εργαστήριο (α) και δευτερευόντως, πιο ανεπίσημα, το ΕΠΤ-ΕΜΠ. Αυτό συνέβη διότι το Εργαστήριο (α) ήταν το παλαιότερο ιστορικά (το τότε Κέντρο Πυρηνικών Ερευνών "Δημόκριτος" ιδρύθηκε το 1959 και το υπόψη Εργαστήριο το 1961), το καλύτερα στελεχωμένο, το περισσότερο γνωστό στο κοινό και το στενότερα συνδεδεμένο με τη διοίκηση της χώρας. Το τελευταίο αυτό χαρακτηριστικό ήταν αποτέλεσμα του ότι δεν είχε ακόμα πλήρως διαμορφωθεί ο ανεξάρτητος ερευνητικός χαρακτήρας του συγκεκριμένου Εθνικού Ερευνητικού Κέντρου, ούτε είχε αναπτυχθεί πλήρως η Ανεξάρτητη Εθνική Αρχή Ακτινοπροστασίας, αυτή που είναι σήμερα γνωστή ως Ελληνική Επιτροπή Ατομικής Ενέργειας. Το αποτέλεσμα αυτής της κατάστασης ήταν ότι το ΕΡΠ πιέσθηκε να εξυπηρετήσει σε πολύ μικρό χρόνο, τεσσάρων με οκτώ εβδομάδων μετά το ατύχημα, τη μέτρηση εκατοντάδων, αν όχι και χιλιάδων δειγμάτων, κυρίως τροφίμων και ιδιαίτερα γάλακτος και τυριών. Για να γίνει κατανοητό το αδύνατο του πράγματος, αναφέρεται εδώ ότι μια πλήρης ανάλυση ενός μόνο δείγματος σε ανιχνευτή Γερμανίου, ακόμα και αν όλα γίνουν όπως πρέπει, απαιτεί από 24 έως 48 ώρες. Από την άλλη, μια προκαταρκτική ανάλυση μικρής διάρκειας ενός δείγματος σε μία ανιχνευτική διάταξη χωρίς τρόπο διαπίστωσης του ποια ισότοπα βρίσκονται στο δείγμα, διαρκεί επίσης αρκετά, δηλαδή από τρία έως πέντε λεπτά. Υπό τις συνθήκες αυτές είναι άξιο θαυμασμού το πώς τελικά το ΕΡΠ κατόρθωσε, μέσα στις αντιξοότητες αυτές, να σχηματίσει, πριν ολοκληρωθεί το καλοκαίρι του 1986, μια αρκετά καλή εικόνα για τη ρύπανση της χώρας εξαιτίας του ατυχήματος τον Απρίλιο (DEMO 86/3 G, 1986). Αναπόφευκτα, η εικόνα αυτή δεν είχε πολύ καλά ποσοτικά χαρακτηριστικά, αλλά οι περιεχόμενες ενδεικτικές πληροφορίες ήταν και πολλές και ικανοποιητικές. Το κοινό χαρακτηριστικό των Εργαστηρίων (β) έως και (ε) ήταν ότι ανήκαν σε ΑΕΙ. Κατά συνέπεια, διατηρώντας την ερευνητική και ακαδημαϊκή τους ανεξαρτησία, δεν δέχθηκαν πίεση από τη διοίκηση ή από το κοινό για μετρήσεις και αναφορές. Σε αυτό έπαιξε ρόλο και το ότι η Πολιτεία δεν είχε επιδιώξει την ένταξή τους σε κάποιο δίκτυο συνεργασίας για την αντιμετώπιση τέτοιων καταστάσεων. Δεδομένης αυτής της ανεξαρτησίας, πρακτικά όλα τα Εργαστήρια αυτού του τύπου επέλεξαν να ασχοληθούν με την αποτύπωση των συνεπειών του ατυχήματος σε πιο εύθετο χρόνο και μετά τη λήξη του μεγάλου θορύβου, ευελιξία που δεν διέθετε ασφαλώς το Εργαστήριο (α).

### 3.2 Ιστορικά στοιχεία για το Εργαστήριο Πυρηνικής Τεχνολογίας ΕΜΠ

Το ΕΠΤ-ΕΜΠ ιδρύθηκε το 1965 και την εποχή του ατυχήματος είχε ήδη συμπληρώσει 20 χρόνια ζωής. Τα πρώτα χρόνια της ζωής του αναπτύχθηκε στο Συγκρότημα Πατησίων στα ονομαζόμενα "Νέα Κτήρια". Περίπου λίγο πριν την εποχή ίδρυσης του Εργαστηρίου, παραχωρήθηκε στο ΕΜΠ έκταση ~1000 στρεμμάτων μεταξύ των συνοικιών Παπάγου και Νέου Ζωγράφου προκειμένου για την πλήρη ανάπτυξη του κτιριακού προγράμματος μιας σύγχρονης Πολυτεχνειούπολης. Η εκκίνηση αυτού του κτιριακού προγράμματος είχε ήδη γίνει από τη δεκαετία του '50 στο από παλαιότερα παραχωρηθέν ΒΔ όμορο τμήμα αυτής της έκτασης, όπου ανεγέρθηκαν το Κτίριο "Υδραυλικής" (σήμερα, 2022, Τομέας Υδατικών Πόρων της Σχολής Πολιτικών Μηχανικών) και το Εργαστήριο Αντοχής Υλικών (σήμερα, 2022, Κτίριο

"Θεοχάρη", Τομέας Μηχανικής της Σχολής Εφαρμοσμένων Μαθηματικών και Φυσικών Επιστημών). Μετά την παραχώρηση και της υπόλοιπης έκτασης, το πρόγραμμα αυτό συνεχίσθηκε με δύο κτίρια: το "Λαμπαδάριο" (σήμερα, 2022, Σχολή Αγρονόμων Τοπογράφων Μηχανικών) προς την πλευρά του Νέου Ζωγράφου και το Κτίριο "Φυσικής" (σήμερα, 2022, Τομέας Φυσικής της Σχολής Εφαρμοσμένων Μαθηματικών και Φυσικών Επιστημών) στο κέντρο της έκτασης. Το Κτίριο "Φυσικής" ολοκληρώθηκε στις αρχές της δεκαετίας του '70 και το πρώτο Εργαστήριο που εγκαταστάθηκε σε αυτό καταλαμβάνοντας μέρος του Ισογείου και του Υπογείου και εμβαδό περίπου 450 $m^2$ ήταν το ΕΠΤ-ΕΜΠ. Κατά το πρώτο διάστημα της εγκατάστασης το Εργαστήριο αντιμετώπισε πολλά προβλήματα. Ενδεικτικά αναφέρονται: (α) παιδικές ασθένειες του κτιρίου, (β) συχνές διακοπές ρεύματος, (γ) ελλιπείς ή κακές τηλεφωνικές συνδέσεις, (δ) προβλήματα ύδρευσης και αποχέτευσης κ.ά. Την εποχή του ατυχήματος, περίπου 15 έτη μετά την εγκατάσταση, δυστυχώς, δεν είχαν λυθεί όλα τα παραπάνω, με κυριότερο πρόβλημα τις διακοπές ρεύματος, παρόλο που η αδιάλειπτη παροχή ηλεκτρικής ισχύος είναι εντελώς απαραίτητη απαίτηση για τη λειτουργία ενός εργαστηρίου αυτού του είδους. Από την άλλη, και μέχρι το 1986, πρέπει να ειπωθεί ότι το ΕΠΤ-ΕΜΠ είχε αποκτήσει, κυρίως με ενέργειες των Καθηγητών Μ.Γ. Αγγελόπουλου και Δ.Ι. Λεωνίδου, αρκετό και ικανό εξοπλισμό που αποδείχθηκε πολύτιμος για τη διερεύνηση των συνεπειών εξαιτίας του ατυχήματος στο Chernobyl. Το ΕΠΤ-ΕΜΠ παρέμεινε στο Κτίριο "Φυσικής" μέχρι το 1999.

### 3.3 Μετρήσεις του Σ.Ε. Σιμόπουλου στο Εργαστήριο Πυρηνικής Τεχνολογίας ΕΜΠ

Το ΕΠΤ-ΕΜΠ, λόγω του ακαδημαϊκού του χαρακτήρα δεν δέχθηκε την πίεση μετρήσεων που υπέστη το ΕΡΠ του Κέντρου Πυρηνικών Ερευνών. Παρόλα αυτά αρκετό πλήθος δειγμάτων κυρίως από τρόφιμα παραδόθηκαν ανεπίσημα και στο ΕΠΤ-ΕΜΠ, για αξιολόγηση. Για τους σκοπούς της αξιολόγησης αυτής χρησιμοποιήθηκαν τόσο ανιχνευτές Ιωδιούχου Νατρίου όσο και ανιχνευτές Γερμανίου μαζί με τον παρελκόμενο εξοπλισμό τους. Σημειώνεται ότι οι προμήθειες αυτού του εξοπλισμού έγιναν ιστορικά ως εξής: (α) Οι ανιχνευτές Ιωδιούχου Νατρίου αποκτήθηκαν με αγορά το έτος 1976. Μία διάταξη με έναν από αυτούς τους ανιχνευτές διακρίνεται στην Εικόνα 4. (β) Το υπολογιστικό σύστημα RT-11, χρήσιμο για τη συλλογή και την ανάλυση φασμάτων -γ φασματοσκοπίας από ανιχνευτές στερεάς κατάστασης αποκτήθηκε με αγορά το έτος 1976 και ένα μέρος του διακρίνεται στην Εικόνα 5. Είναι σε λειτουργική κατάσταση αλλά δεν έχει χρησιμοποιηθεί μετά το έτος 2000 λόγω αντικατάστασής του με νεότερα συστήματα. (γ) Ο ανιχνευτής τύπου Ge(Li) αποκτήθηκε με αγορά από τον οίκο CANBERRA των ΗΠΑ, το έτος 1978. Ο ανιχνευτής αυτός λειτούργησε αδιάλειπτα για περίπου 30 έτη, διάστημα που αποτελεί μάλλον αξεπέραστη επίδοση για ανιχνευτή αυτής της τεχνολογίας. Σήμερα φυλάσσεται στο ΕΠΤ-ΕΜΠ για να θυμίζει το παρελθόν και διακρίνεται στην Εικόνα 6.

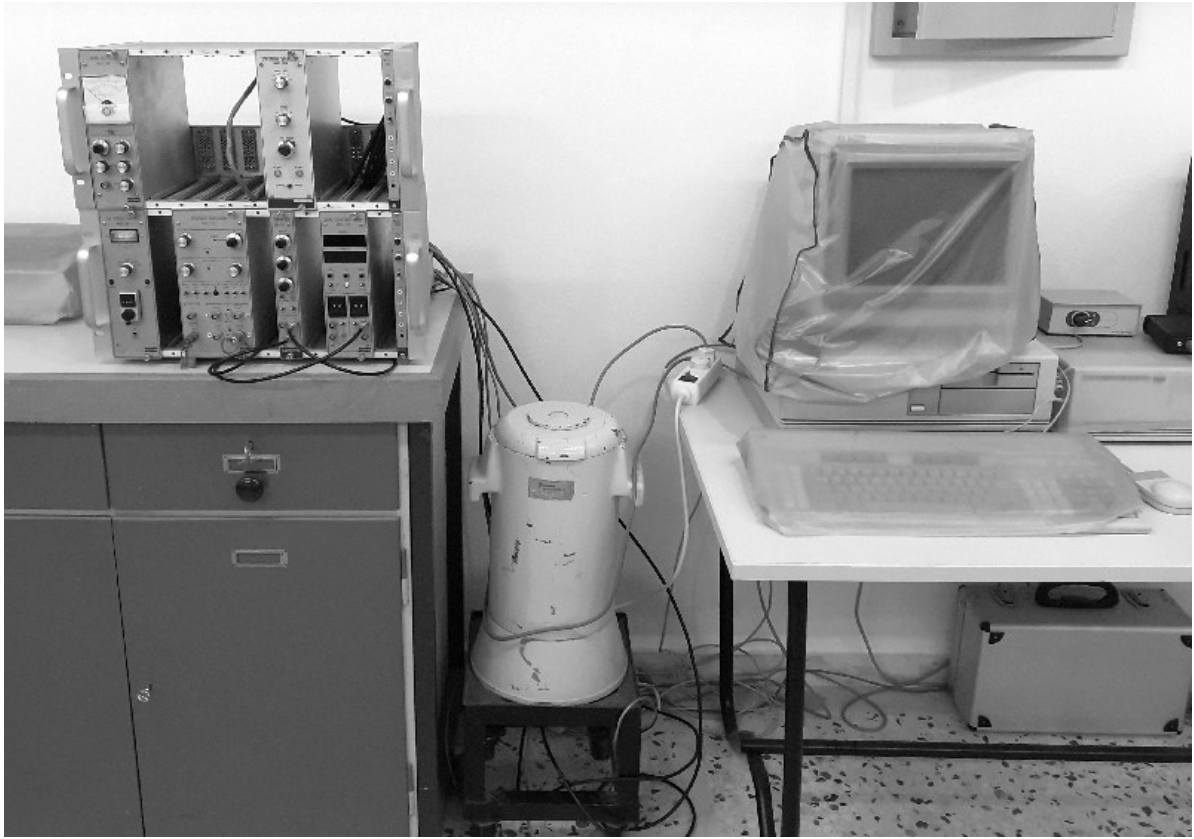

Εικόνα 4. Διάταξη με ανιχνευτή Ιωδιούχου Νατρίου από αυτές που χρησιμοποιήθηκαν στην ερευνητική δουλειά του Σ.Ε. Σιμόπουλου το 1986, όπως είναι και λειτουργεί σήμερα (2022)

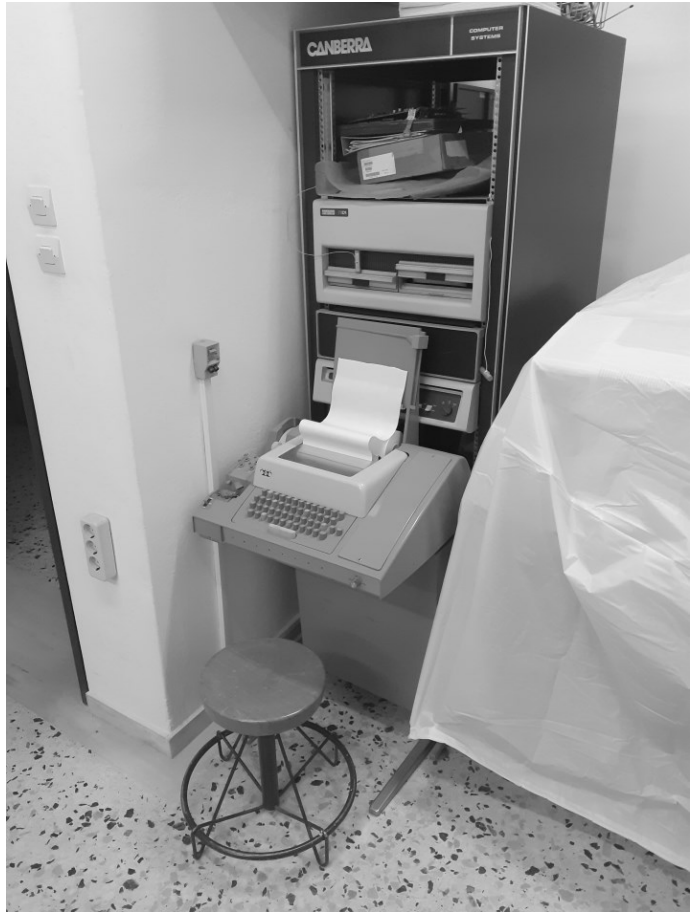

Εικόνα 5. Το υπολογιστικό σύστημα RT-11 της εταιρείας Digital, όπως υπάρχει σήμερα (2022)

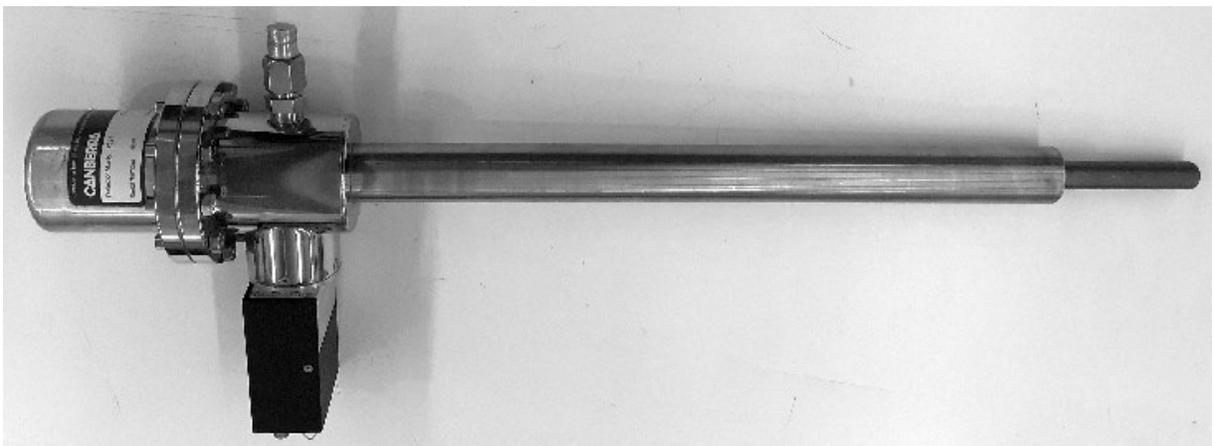

Εικόνα 6. Ο ανιχνευτής τύπου Ge(Li) του οίκου CANBERRA - ΗΠΑ, όπως φυλάσσεται σήμερα

Στην Εικόνα 7α φαίνεται η θωράκιση (ιδιοκατασκευή του ΕΠΤ-ΕΜΠ), στην οποία είχε τοποθετηθεί. (δ) Ο ανιχνευτής τύπου υπερκαθαρού Γερμανίου αποκτήθηκε με αγορά επίσης από τον οίκο CANBERRA των ΗΠΑ το έτος 1982. Ο ανιχνευτής αυτός λειτουργεί ακόμα φθάνοντας σήμερα (2022) τα 40 έτη ζωής και φαίνεται στην Εικόνα 7β.

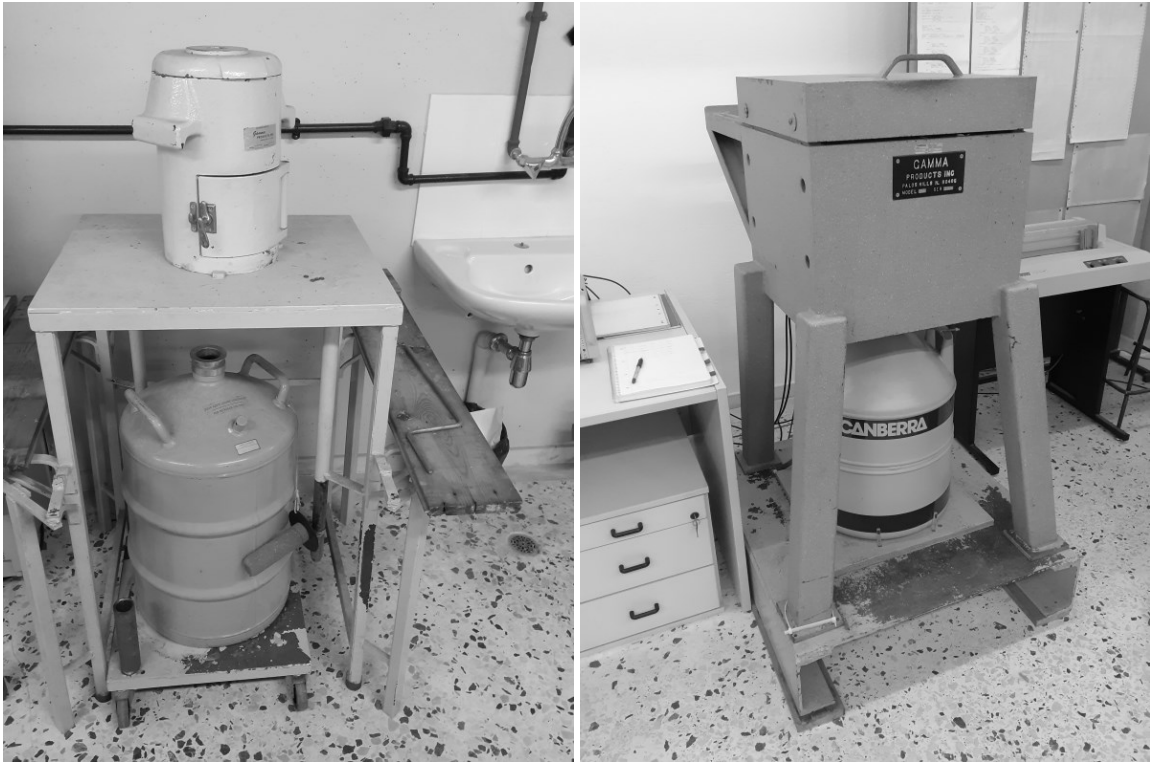

Εικόνα 7. (α) Αριστερά: Η θωράκιση του ανιχνευτή της Εικόνας 5.
(β) Δεξιά: Ο ανιχνευτής τύπου Υπερκαθαρού Γερμανίου του οίκου CANBERRA - ΗΠΑ,
όπως λειτουργεί σήμερα (2022)

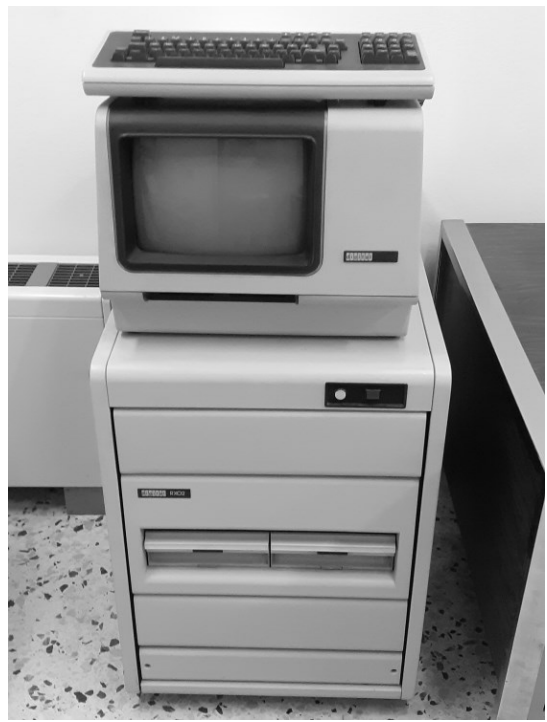

Εικόνα 8. Το υπολογιστικό σύστημα LSI-11 της εταιρείας Digital, όπως υπάρχει σήμερα (2022)

Τέλος, (ε), το υπολογιστικό σύστημα LSI-11, εξίσου χρήσιμο για τη συλλογή και την ανάλυση φασμάτων -γ φασματοσκοπίας από ανιχνευτές στερεάς κατάστασης αποκτήθηκε με δωρεά από την τότε εταιρεία Digital ΗΠΑ το έτος 1982 και ένα μέρος του διακρίνεται στην Εικόνα 8. Είναι σε λειτουργική κατάσταση αλλά δεν έχει χρησιμοποιηθεί μετά το έτος 2000 λόγω αντικατάστασής του με νεότερα συστήματα. Ανύποπτο χρόνο πριν το ατύχημα στο Chernobyl, τον εξοπλισμό αυτό συνέθεσε σε ενιαία συνεκτική και λειτουργική ενότητα διαθέσιμη και σε ετοιμότητα για μετρήσεις ο Σ.Ε. Σιμόπουλος από κοινού με το μέλος ΕΕΠ του ΕΠΤ-ΕΜΠ Μηχανολόγο - Ηλεκτρολόγο Μηχανικό Δ. Πετρόπουλο.

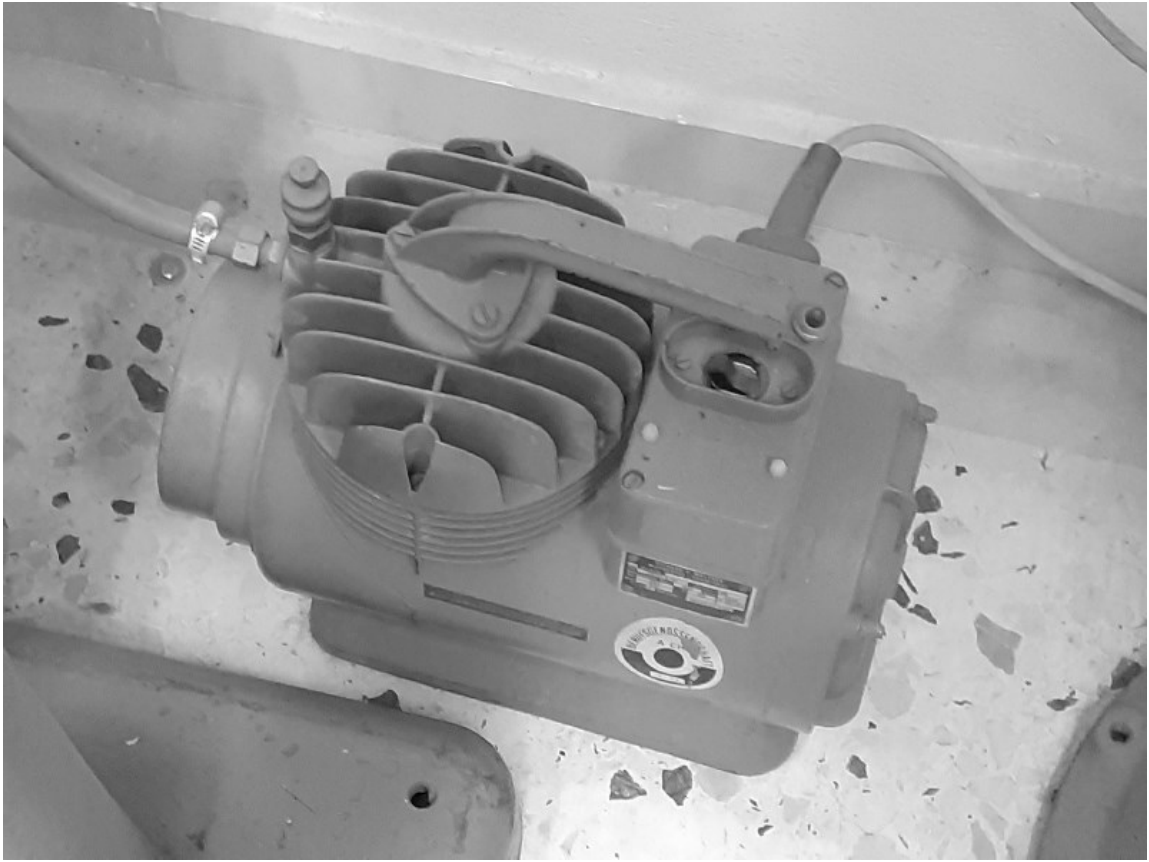

Εικόνα 9. Αντλία που χρησιμοποιήθηκε για τις πρώτες δειγματοληψίες ατμοσφαιρικού αέρα επί φίλτρων από το ΕΠΤ-ΕΜΠ στη περιοχή της Αθήνας, τον Μάιο του 1986

Κατά τις πρώτες ημέρες μετά το ατύχημα πραγματοποιήθηκαν μετρήσεις ατμοσφαιρικού αέρα με τη βοήθεια αντλίας αναρρόφησης και φίλτρων που παραχώρησε στο ΕΠΤ-ΕΜΠ ο Καθηγητής της Σχολής Μηχανολόγων Μηχανικών ΕΜΠ Δ. Κουρεμένος. Η αντλία αυτή υπάρχει ακόμα, λειτουργεί και διακρίνεται στην Εικόνα 9.

Στον αέρα του περιβάλλοντος που πέρασε με τη βοήθεια της αντλίας αυτής από το φίλτρο της Εικόνας 10α ανιχνεύθηκε Am-241 (δηλ. απειροστό, φυσικά, τμήμα από την καρδιά του πυρηνικού αντιδραστήρα στο Chernobyl), διαπίστωση που έδειξε και τη σοβαρότητα του ατυχήματος.

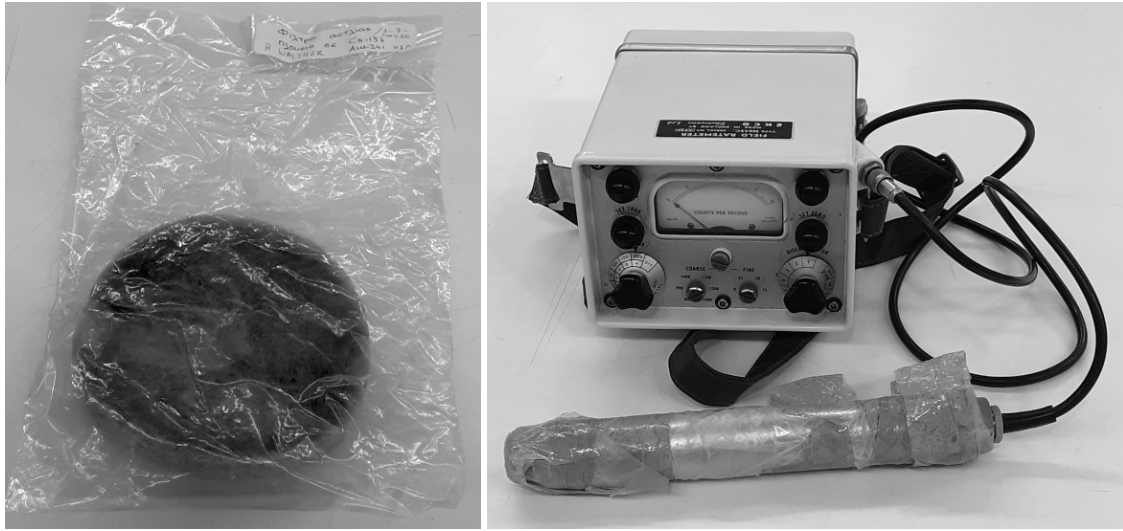

Εικόνα 10. (α) Αριστερά: Φίλτρο στο οποίο βρέθηκε Am-241 τον Μάιο του 1986
(β) Δεξιά: Παλαιός φορητός ανιχνευτής Ιωδιούχου Νατρίου, οίκου ECKO που χρησιμοποιήθηκε για τον εντοπισμό περιοχών με αυξημένη εναπόθεση Cs-137

Μετά την ένταση των πρώτων ημερών μετά το ατύχημα, ο Σ.Ε. Σιμόπουλος συνειδητοποίησε ότι επείγει η αντιπροσωπευτική δειγματοληψία των επιφανειακών εδαφών στην Ελλάδα για την αποτύπωση της εναπόθεσης ισοτόπων εξαιτίας του ατυχήματος πριν έρθουν οι βροχές του φθινοπώρου και του χειμώνα που θα αλλοίωναν τις αποδείξεις. Ταυτόχρονα συνειδητοποίησε ότι εκτός από τη συλλογή επείγει και η μέτρηση των δειγμάτων ώστε, αν κάποια βρεθούν με υψηλές συγκεντρώσεις σε ισότοπα, να ενημερωθεί έγκαιρα και με ακρίβεια η Πολιτεία, προκειμένου να ληφθούν, αν χρειάζονταν, τα αναγκαία μέτρα. Με βάση αυτές τις απαιτήσεις, ο Σ.Ε. Σιμόπουλος εργάσθηκε πρώτα και για ένα περίπου δίμηνο, για να εξασφαλίσει μια μέθοδο σύντομης και αξιόπιστης μέτρησης της συγκέντρωσης Cs-137. Αυτή βρέθηκε και υλοποιήθηκε όπως περιγράφεται στο Simopoulos (1989). Έπειτα και μετά τη λήξη των εκπαιδευτικών του υποχρεώσεων στο ΕΜΠ ξεκίνησε, το θέρος του 1986, τη δειγματοληψία επιφανειακών εδαφών ως εξής: (α) χωρίς καμία οικονομική ενίσχυση, (β) διαθέτοντας μέχρι καταστροφής το προσωπικό του αυτοκίνητο (ένα Audi 80) και (γ) διαθέτοντας απροσδιόριστες ώρες από τον προσωπικό του χρόνο. Μια πρόχειρη ανάλυση του Πίνακα 1 και του Πίνακα 2 στο Simopoulos (1989) δείχνει ότι μεταφέρθηκαν στο ΕΠΤ-ΕΜΠ, σε πέντε δόσεις, περίπου δύο τόνοι, μαζί με τη φύρα, δειγμάτων επιφανειακού εδάφους σχεδόν από όλη την ηπειρωτική χώρα. Προκύπτει επίσης ότι πραγματοποιήθηκαν τουλάχιστον 10000 km διαδρομών με αυτοκίνητο και ένας σημαντικός αριθμός διανυκτερεύσεων εκτός έδρας. Για την επί τόπου προκαταρκτική διαπίστωση τυχόν αυξημένων συγκεντρώσεων Cs-137 στα εδάφη πριν τη δειγματοληψία χρησιμοποιήθηκε ένας παλαιός φορητός ανιχνευτής Ιωδιούχου Νατρίου, ο οποίος και σήμερα λειτουργεί και διακρίνεται στην Εικόνα 10β. Τα συλλεχθέντα το 1986 1500 δείγματα διατηρούνται σήμερα στο πλαστικό κυλινδρικό τους δοχείο, στο οποίο μετρήθηκαν, στην αποθήκη δειγμάτων του ΕΠΤ-ΕΜΠ ως αποδεικτικά στοιχεία. Το μέρος της αποθήκης με τα δείγματα αυτά διακρίνεται στην Εικόνα 11.

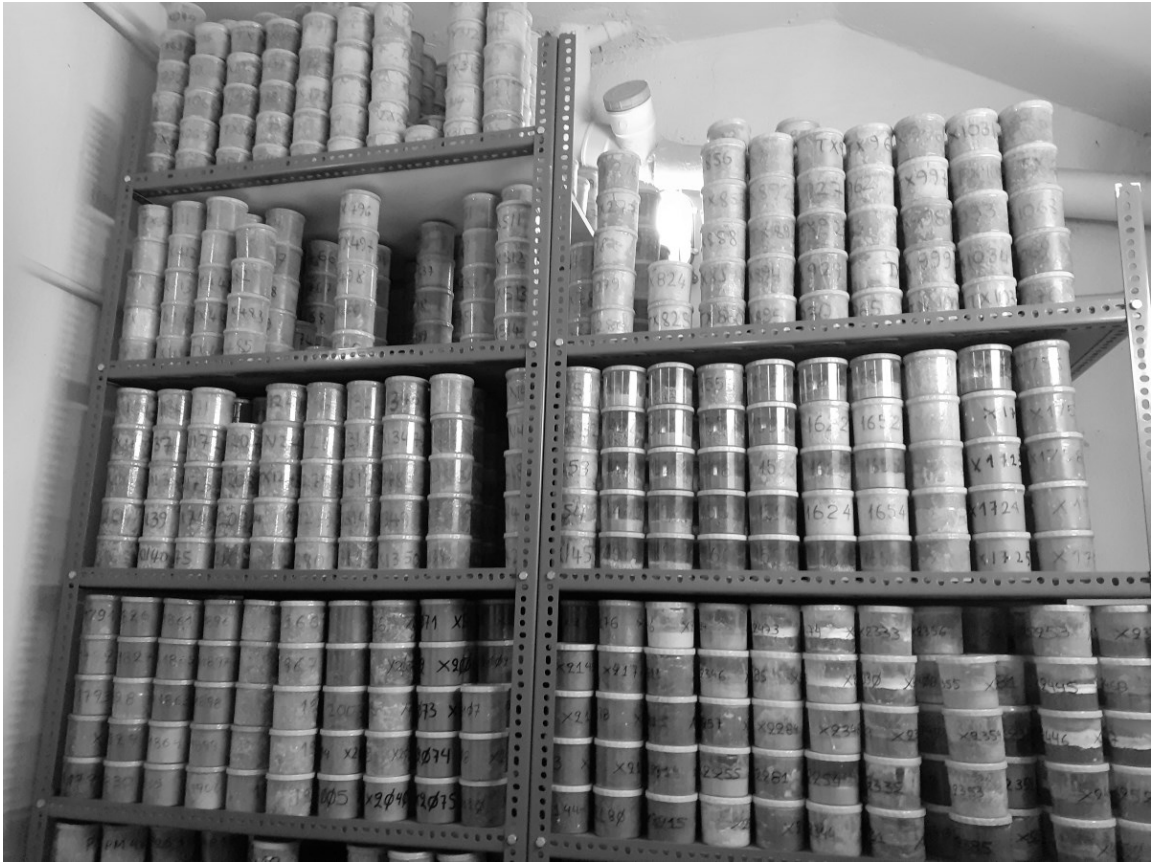

Εικόνα 11. Η αποθήκη με τα δείγματα επιφανειακών εδαφών της χώρας που συλλέχθηκαν το 1986 (περίπου 1500) και μεταξύ 1987 - 2007 (περίπου 1000 επιπλέον)

Φυσικά τα σχετικά δείγματα είναι σήμερα πολύ περισσότερα, γύρω στα 2500, διότι οι δειγματοληψίες συνεχίσθηκαν από τον Σ.Ε. Σιμόπουλο και τους συνεργάτες του για τα επόμενα περίπου 20 έτη. Τα αποτελέσματα των μετρήσεων του 1986 αποτυπώθηκαν σε πρόχειρους ιδιόχειρους χάρτες που έφτιαξε ο ίδιος ο Σ.Ε. Σιμόπουλος. Ένα χαρακτηριστικό αντίγραφο με μετρήσεις κατά Νομό και με χειρόγραφες σημειώσεις του ιδίου δίνεται στην Εικόνα 12.

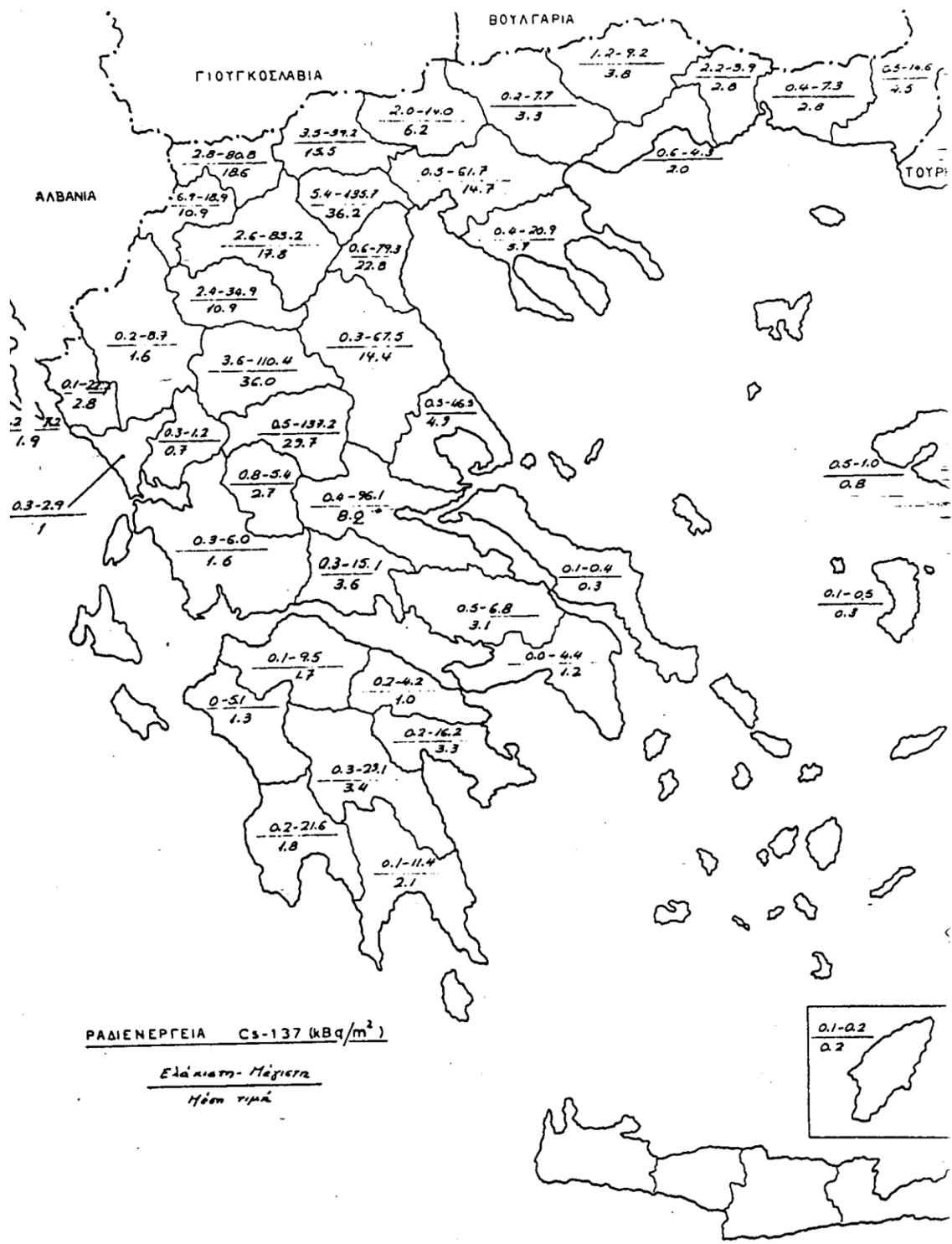

Εικόνα 12. Ραδιενέργεια Cs-137 (σε kBqm$^{-2}$) ανά Νομό της Ελλάδας. Ιδιόχειρος χάρτης από τις εκθέ-
σεις - αναφορές του Σ.Ε. Σιμόπουλου προς τη διοίκηση του ΕΜΠ
(Σιμόπουλος 1986, Σιμόπουλος 1987)

Ένα δεύτερος χάρτης με μέγιστες τιμές συγκεντρώσεων Cs-137 στα επιφανειακά εδάφη, επίσης ιδιόχειρος, δίνεται στην Εικόνα 13.

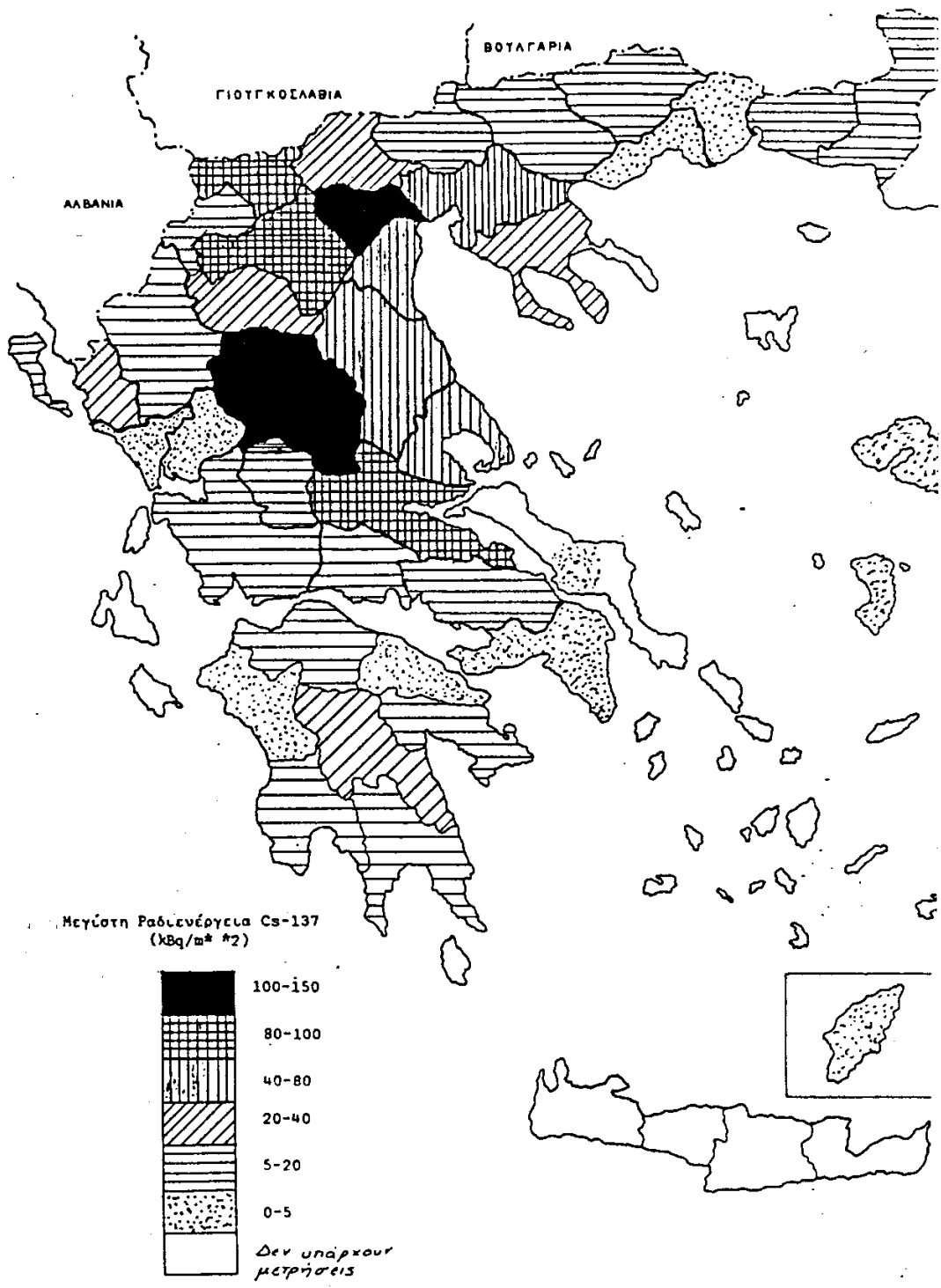

Εικόνα 13. Μέγιστη ραδιενέργεια Cs-137 (σε kBqm$^{-2}$) ανά Νομό της Ελλάδας. Ιδιόχειρος χάρτης από τις εκθέσεις - αναφορές του Σ.Ε. Σιμόπουλου προς τη διοίκηση του ΕΜΠ
(Σιμόπουλος 1986, Σιμόπουλος 1987)

Επισημαίνεται ότι οι εκθέσεις- αναφορές με τα αποτελέσματα των μετρήσεων που οδήγησαν στους πρόχειρους χάρτες της Εικόνας 12 και της Εικόνας 13, διακινήθηκαν τόσο εσωτερικά στο ΕΜΠ, όσο και ευρύτερα προς άλλους Δημόσιους Οργανισμούς και τα λοιπά όργανα της Πολιτείας μέσω εμπιστευτικής διαδικασίας (Σιμόπουλος 1986, Σιμόπουλος 1987) . Ο λόγος ήταν για να αποφευχθεί τυχόν ανησυχία στις περισσότερο ρυπασμένες περιοχές της Καρδίτσας και της Νάουσας, καθόσον δεν είχαν ακόμη ολοκληρωθεί οι μελέτες που έδειξαν ότι ακόμα και αυτή η υψηλότερη ρύπανση, είχε αμελητέες επιπτώσεις σε βιολογικό επίπεδο (π.χ. Γεροντίδου, 1994).Οι χάρτες αυτοί υστερούν σε ποιότητα και ανάλυση από άλλους που μπορούν να παράγονται με προηγμένες μεθόδους με βάση τα ίδια αποτελέσματα μετρήσεων. Για αυτό και το ΕΠΤ-ΕΜΠ επανήλθε σε αυτούς παράγοντας αυτόν της Εικόνας 14.

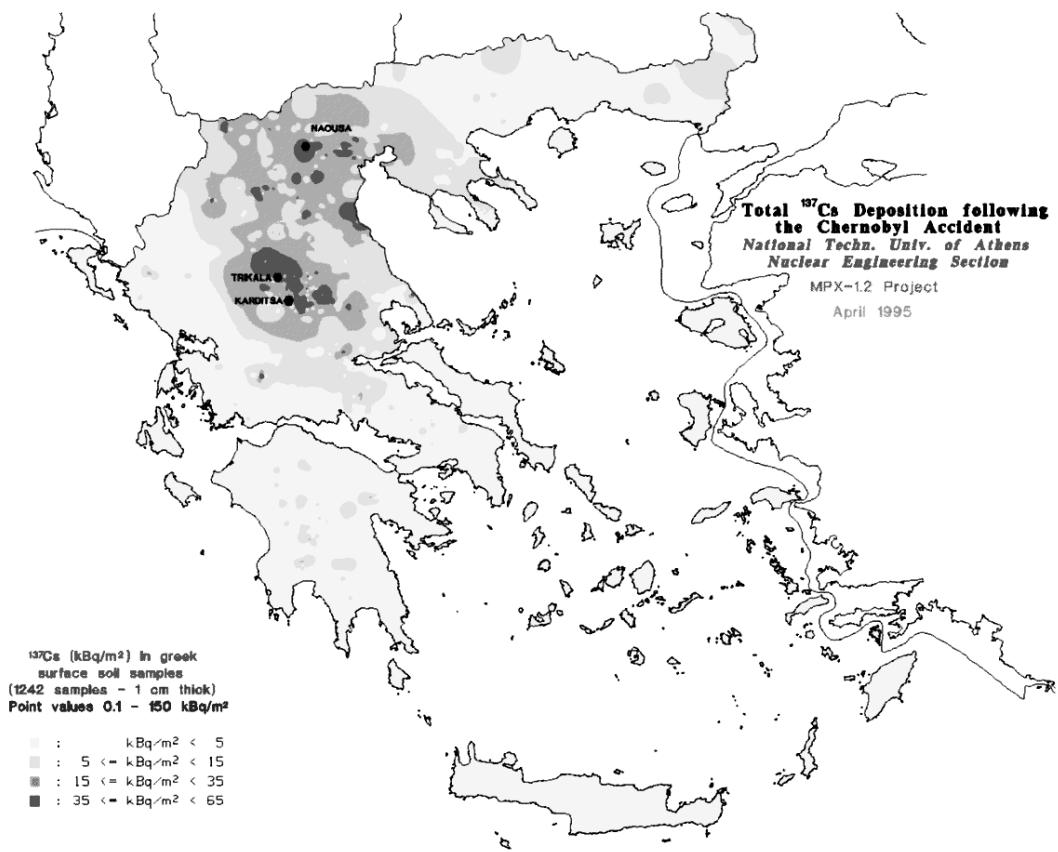

Εικόνα 14. Χάρτης απόθεσης Cs-137 (σε $kBqm^{-2}$) στα επιφανειακά εδάφη της Ελλάδας λόγω του ατυχήματος στο Chernobyl (Petropoulos, 2001)

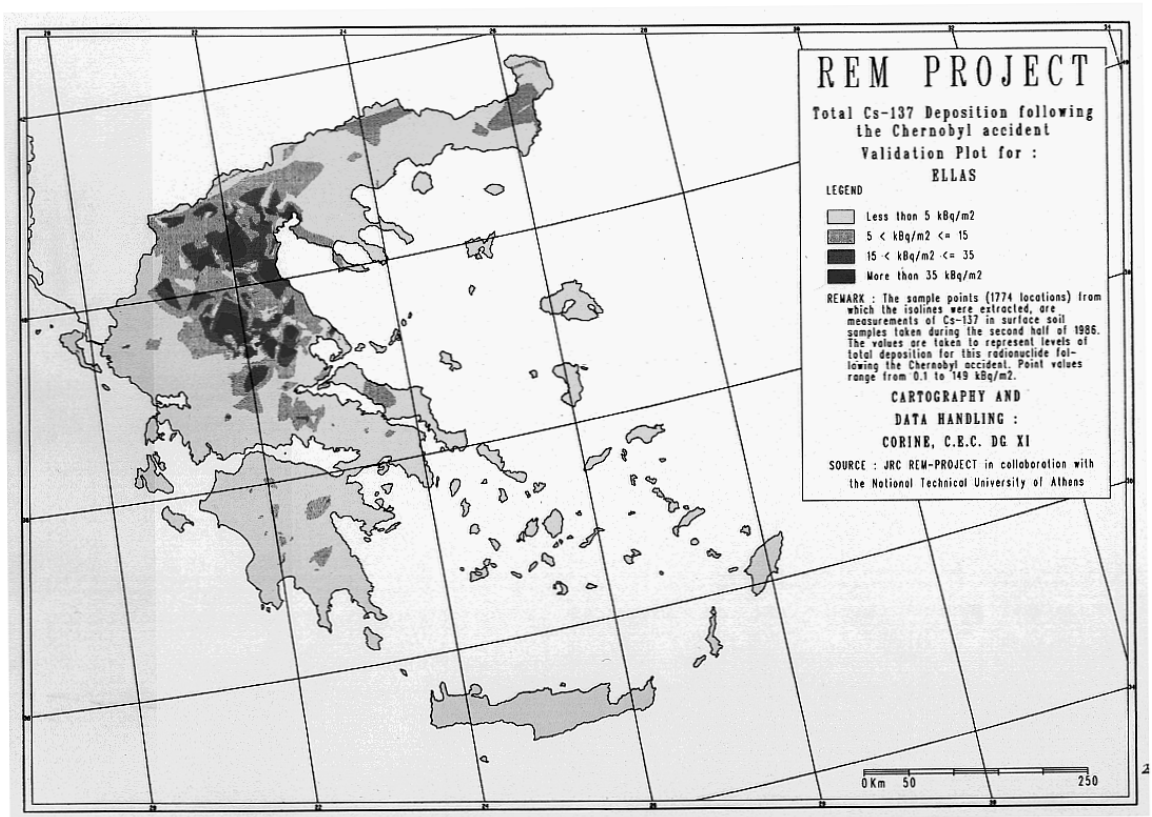

Εικόνα 15. Χάρτης απόθεσης Cs-137 (σε kBqm$^{-2}$) στα επιφανειακά εδάφη της Ελλάδας λόγω του ατυχήματος στο Chernobyl με βάση τα ίδια πειραματικά αποτελέσματα, από τα οποία προέκυψε ο χάρτης της Εικόνας 14. Παρουσιάσθηκε ως εξαιρετικό παράδειγμα πιστής χαρτογράφησης με βάση δεδομένα ρύπανσης εδάφους στη Σύνοδο του Ρίο για το κλίμα (1992). Ο τελικός χάρτης δημοσιεύθηκε από την ερευνητική ομάδα De Cort (1998)

Ο χάρτης της Εικόνας 14 είναι πρακτικά όμοιος με αυτόν της Εικόνας 15, ο οποίος ετοιμάσθηκε με βάση τα ίδια αποτελέσματα από διεθνή ερευνητική ομάδα που εργάσθηκε για τη λεπτομερή χαρτογράφηση της απόθεσης του Cs-137 σε όλη την Ευρώπη και την Ευρωπαϊκή Ρωσία (De Cort, 1998). Η ομάδα αυτή παρουσίασε στη Σύνοδο του Ρίο για το κλίμα (1992) τον χάρτη της Εικόνας 15 ως εξαιρετικό παράδειγμα πιστής χαρτογράφησης με βάση δεδομένα ρύπανσης εδάφους.

### 4. Θετικά αποτελέσματα του ατυχήματος

Η Ελληνική Πολιτεία, μετά από τα όσα συνέβησαν το 1986 και τη ρύπανση που προκλήθηκε στην Ελλάδα από το ατύχημα στο Chernobyl, όπως αποτυπώθηκε στο Simopoulos (1989) αλλά και αλλού, έλαβε, με αργό βήμα είναι η αλήθεια, πολύ ουσιαστικά μέτρα για τη βελτίωση της απόκρισης της χώρας σε τέτοιες ή παρόμοιες καταστάσεις. Τα μέτρα αυτά δίνονται ακολούθως χωρίς απαραίτητα να τηρείται η χρονολογική σειρά:

(α) Επανιδρύθηκε η Ελληνική Επιτροπή Ατομικής Ενέργειας (ΕΕΑΕ) ως ανεξάρτητη υπηρεσία υπαγόμενη στη Γενική Γραμματεία Έρευνας.

(β) Ανανεώθηκε η νομοθεσία που αφορά στα Εθνικά Ερευνητικά Κέντρα της χώρας και ενθαρρύνθηκε θεσμικά η ανεξαρτησία τους από τη διοίκηση.

(γ) Αναπτύχθηκε, με ευθύνη της ΕΕΑΕ, δίκτυο κατάλληλων ανιχνευτών γεωγραφικά κατανεμημένων σε όλη τη χώρα, ικανών να μεταδίδουν σήματα σε πραγματικό χρόνο σχετικά με τα επίπεδα ραδιενέργειας περιβάλλοντος. Στόχος του δικτύου είναι να εντοπίζει αδικαιολόγητες αυξήσεις που δυνατόν να συνδέονται με πιθανή διαρροή ραδιενεργών ρύπων από γειτονική χώρα που διαθέτει πυρηνικούς αντιδραστήρες.

(δ) Αναπτύχθηκε, με ευθύνη της ΕΕΑΕ, δίκτυο συνεργαζομένων Ελληνικών Εργαστηρίων από αυτά που υπάρχουν κυρίως στα ΑΕΙ, το οποίο να μπορεί να συνδράμει την ΕΕΑΕ σε μετρήσεις και εκτιμήσεις ρύπανσης. Ανάμεσα στα εργαστήρια αυτά βρίσκεται φυσικά και το ΕΠΤ-ΕΜΠ.

(ε) Δόθηκαν εφάπαξ χρηματοδοτήσεις μετρίου ύψους σε ορισμένα από τα συνεργαζόμενα εργαστήρια προκειμένου αυτά να αντιμετωπίσουν τις άμεσες ανάγκες τους σε βάθος πενταετίας από το 1986 έως και το 1990. Το ΕΠΤ-ΕΜΠ έλαβε το αναγκαίο ποσό για την αναβάθμιση των υπολογιστικών του συστημάτων. Στο πλαίσιο αυτό το ΕΠΤ-ΕΜΠ εγκατέστησε ένα από τα πρώτα, αν όχι μάλιστα το πρώτο UNIX σύστημα του ΕΜΠ, το 1987. Το σύστημα αυτό, ένας supermini computer οίκου Hewlett Packard τύπου HP-9000/320 με λειτουργικό σύστημα HP-UX έκδοσης 7.03 (UNIX System V), το οποίο διακρίνεται στην Εικόνα 16, βρίσκεται σε λειτουργική κατάσταση αλλά δεν έχει χρησιμοποιηθεί μετά το έτος 2008 λόγω αντικατάστασής του με νεότερα συστήματα.

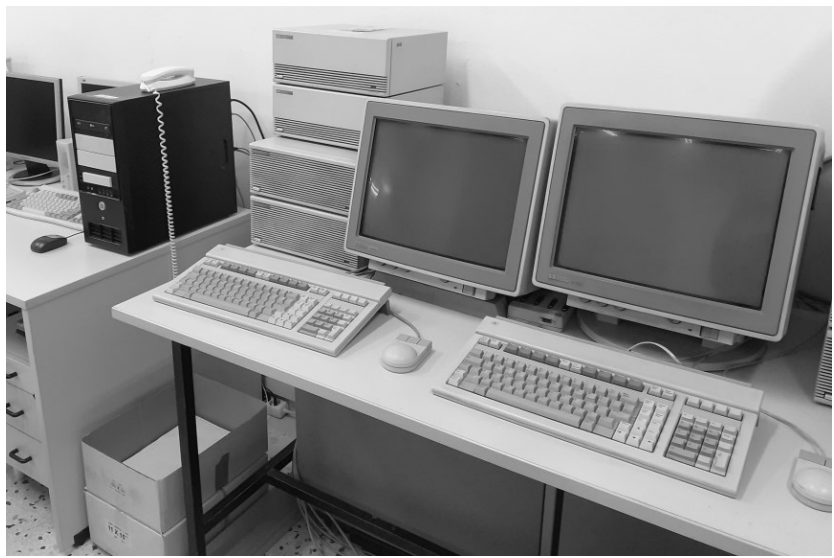

Εικόνα 16. Ο supermini υπολογιστής τύπου HP-9000/320 με λειτουργικό σύστημα HP-UX 7.03 όπως βρίσκεται στο ΕΠΤ-ΕΜΠ σήμερα (2022). Έτος αγοράς 1987.

και τέλος

(στ) Ενισχύθηκε η ΕΕΑΕ, τόσο με εξοπλισμό, όσο και με την πρόσληψη εξειδικευμένων επιστημόνων, οι οποίοι στελέχωσαν με κατάλληλο και επαρκή τρόπο το νέο Τμήμα Ελέγχου Ραδιενέργειας Περιβάλλοντος της Διεύθυνσης Αδειών και Ελέγχων.

### 5. Πρόσωπα που συνέβαλαν στις μετρήσεις στο ΕΠΤ-ΕΜΠ

Σε αυτήν την ενότητα της ιστορίας πρέπει πρώτα να γίνει αναφορά στον εμβληματικό Καθηγητή του ΕΜΠ Μιχαήλ Γ. Αγγελόπουλο[†]. Ο Μ.Γ. Αγγελόπουλος γεννήθηκε στην Αθήνα το 1926. Αποφοίτησε από τη Σχολή Μηχανολόγων - Ηλεκτρολόγων Μηχανικών ΕΜΠ. Έκανε μεταπτυχιακές σπουδές στη Δυτική Γερμανία (στην περιοχή των Υψηλών Τάσεων) και στο Ηνωμένο Βασίλειο (στην Πυρηνική Τεχνολογία). Κατά τη θητεία του ως έκτακτος Καθηγητής Πυρηνικής Τεχνολογίας ιδρύθηκε το Εργαστήριο Πυρηνικής Τεχνολογίας του ΕΜΠ. Εκλέχθηκε Τακτικός Καθηγητής του Ιδρύματος στη νέα Έδρα Πυρηνικής Τεχνολογίας το 1968. Σύμφωνα με το λαϊκό μύθο που μας μεταφέρει ο πρώην διοικητής της ΔΕΗ Ραφαήλ Μωυσής, οι συμφοιτητές του τον θεωρούσαν ένα από τα τρία ιερά τέρατα ιδιοφυίας των Μηχανολόγων - Ηλεκτρολόγων Μηχανικών του ΕΜΠ, οι άλλοι δύο ήταν ο Ηλίας Γυφτόπουλος και ο Δημήτρης Συμεών. Ανέλαβε διοικητής της ΔΕΗ, μετά τη μεταπολίτευση και μέχρι το 1979. Ως επιστήμονας ήταν πάντοτε υπέρ των πυρηνικών αντιδραστήρων μικρής ισχύος διότι πίστευε ότι, σε περίπτωση ατυχήματος, αυτό θα ήταν καλύτερα διαχειρίσιμο. Επιπλέον, έκρινε ότι οι αντιδραστήρες μικρής ισχύος μπορούν να προστατεύονται καλύτερα από εξωτερικούς βλαπτικούς παράγοντες. Το ατύχημα στο Chernobyl τον βρήκε στις επάλξεις της διδασκαλίας Πυρηνικής Τεχνολογίας, με έμφαση στους αντιδραστήρες, στη Σχολή Μηχανολόγων Μηχανικών. Στήριξε με όλες του τις δυνάμεις την προσπάθεια του Σ.Ε. Σιμόπουλου σχετικά με την αποτύπωση της ρύπανσης της Ελλάδας, όσο το δυνατόν πιο σύντομα και όσο το δυνατό πιο αποδοτικά. Αφυπηρέτησε το 1994. Στην Εικόνα 17 διακρίνεται ο πρωθυπουργός Κ. Καραμανλής μαζί με τον Μ.Γ. Αγγελόπουλο.

Δεύτερο σημαντικό πρόσωπο της ίδιας ιστορίας είναι ο Καθηγητής Πυρηνικής Τεχνολογίας Δρ. Δημήτριος Ι. Λεωνίδου. Ο Δ.Ι. Λεωνίδου γεννήθηκε το 1937 στη Χίο. Αποφοίτησε ως φυσικός από το ΕΚΠΑ. Έκανε το διδακτορικό του στην Πυρηνική Τεχνολογία στο Ηνωμένο Βασίλειο. Εκλέχθηκε Μόνιμος Καθηγητής της Β' Έδρας Πυρηνικής Τεχνολογίας του ΕΜΠ το 1981. Το ατύχημα στο Chernobyl τον βρήκε στις επάλξεις της διδασκαλίας εξειδικευμένων μαθημάτων σχετικών με την Πυρηνική Τεχνολογία και τις εφαρμογές της με έμφαση στην Πειραματική Πυρηνική Τεχνολογία και τη Δοσιμετρία και την Ακτινοπροστασία. Σε αυτόν οφείλονται οι αγορές τόσο του ανιχνευτή Ge(Li), όσο και του υπολογιστικού συστήματος RT-11, καθώς και άλλου περιφερειακού εξοπλισμού που αποδείχθηκαν πολύτιμα για την διερεύνηση των συνεπειών του ατυχήματος. Αφυπηρέτησε το 2004. Στην Εικόνα 18, αριστερά διακρίνεται ο Καθηγητής Δ.Ι. Λεωνίδου το 2002.

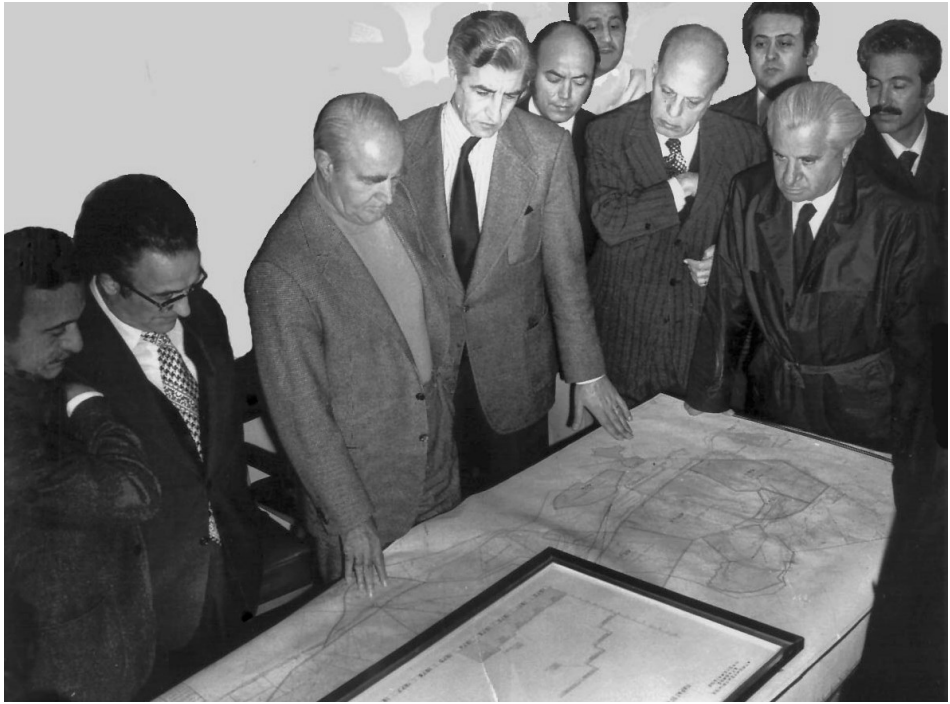

Εικόνα 17. Ο Καθηγητής Μ.Γ. Αγγελόπουλος με τον πρωθυπουργό της χώρας Κ. Καραμανλή (και οι δύο με τα χέρια στο τραπέζι) στο συγκρότημα των ΑΗΣ Καρδιάς το 1975 (Ζαραφίδης και Παυλουδάκης, 2008)

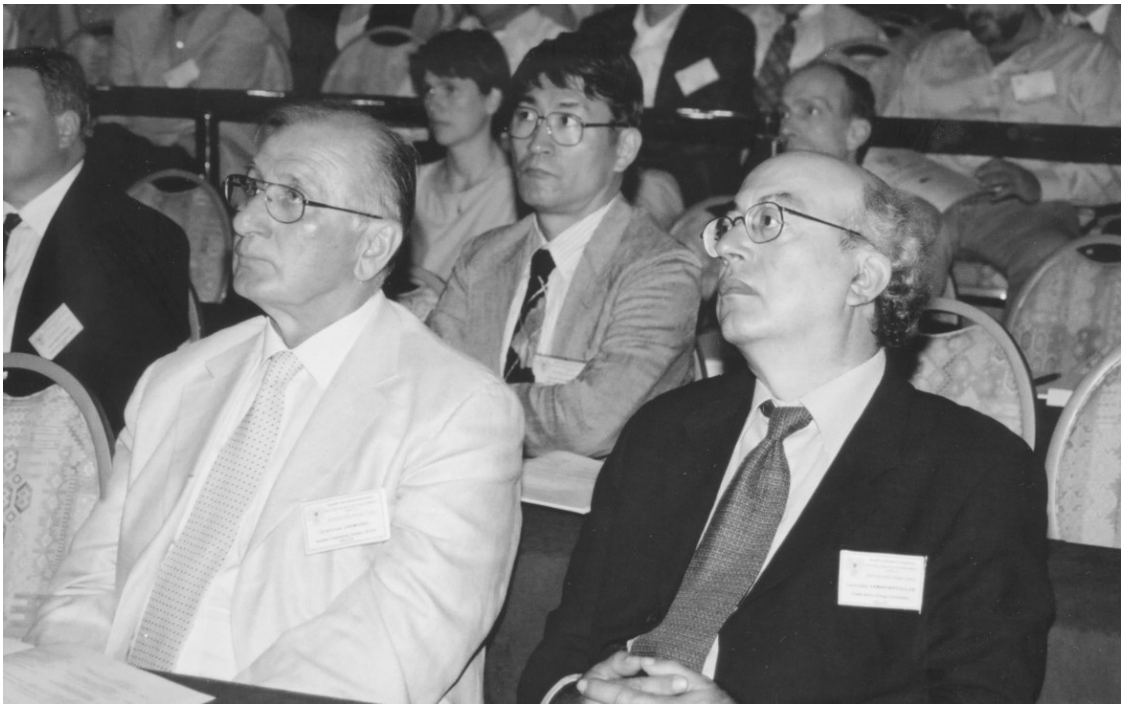

Εικόνα 18. Ο Καθηγητής Δ.Ι. Λεωνίδου (αριστερά) παρακολουθεί τις εργασίες του Διεθνούς Συνεδρίου Natural Radiation Environment VII, στη Ρόδο, τον Μάιο του 2002. Δεξιά ο τότε Πρόεδρος της ΕΕΑΕ Καθηγητής (ΠΑΠΕΙ) Λεωνίδας Καμαρινόπουλος

Από τα υπόλοιπα στελέχη του ΕΠΤ-ΕΜΠ την περίοδο του ατυχήματος στο Chernobyl, αξίζει ασφαλώς να μνημονευτούν το μέλος ΕΕΠ, Δημήτρης Πετρόπουλος, Μηχανολόγος - Ηλεκτρολόγος Μηχανικός ΕΜΠ, το μέλος ΕΕΠ, Νίκος Δημητρακόπουλος†, Φυσικός ΕΚΠΑ, και το μέλος ΕΔΤΠ-ΥΕ, Βασίλης Ηλίας†, πιστοποιημένος τεχνικός ηλεκτρολόγος μεγάλης ισχύος. Δυστυχώς, για τον Δ. Πετρόπουλο, παρόλο που η συμβολή του στη μέθοδο των μετρήσεων είναι καλά αναγνωρισμένη και μέρος της προφορικής ιστορίας του ΕΠΤ-ΕΜΠ, δεν στάθηκε δυνατό να βρεθούν επιπλέον στοιχεία στο αρχείο του ΕΠΤ-ΕΜΠ. Στην Εικόνα 19 αριστερά, διακρίνεται ο Ν. Δημητρακόπουλος. Στην Εικόνα 19 δεξιά διακρίνεται ο Β. Ηλίας. Πρέπει να σημειωθεί ότι μεγάλο μέρος της προετοιμασίας των συλλεχθέντων δειγμάτων εδάφους για μέτρηση, πέρασε από τα χέρια του Β. Ηλία, ο οποίος εργαζόμενος με συστηματικό τρόπο πέτυχε τη διαρκή τροφοδοσία των μετρητικών διατάξεων με δείγματα έτσι ώστε να μην χάνεται σχεδόν καθόλου χρόνος. Ο Ν. Δημητρακόπουλος υπήρξε ένας άοκνος και σιωπηλός εργάτης της εργαστηριακής εκπαίδευσης και των φροντιστηρίων για τους φοιτητές που επέλεγαν μαθήματα Πυρηνικής Τεχνολογίας στη Σχολή Μηχανολόγων ΕΜΠ.

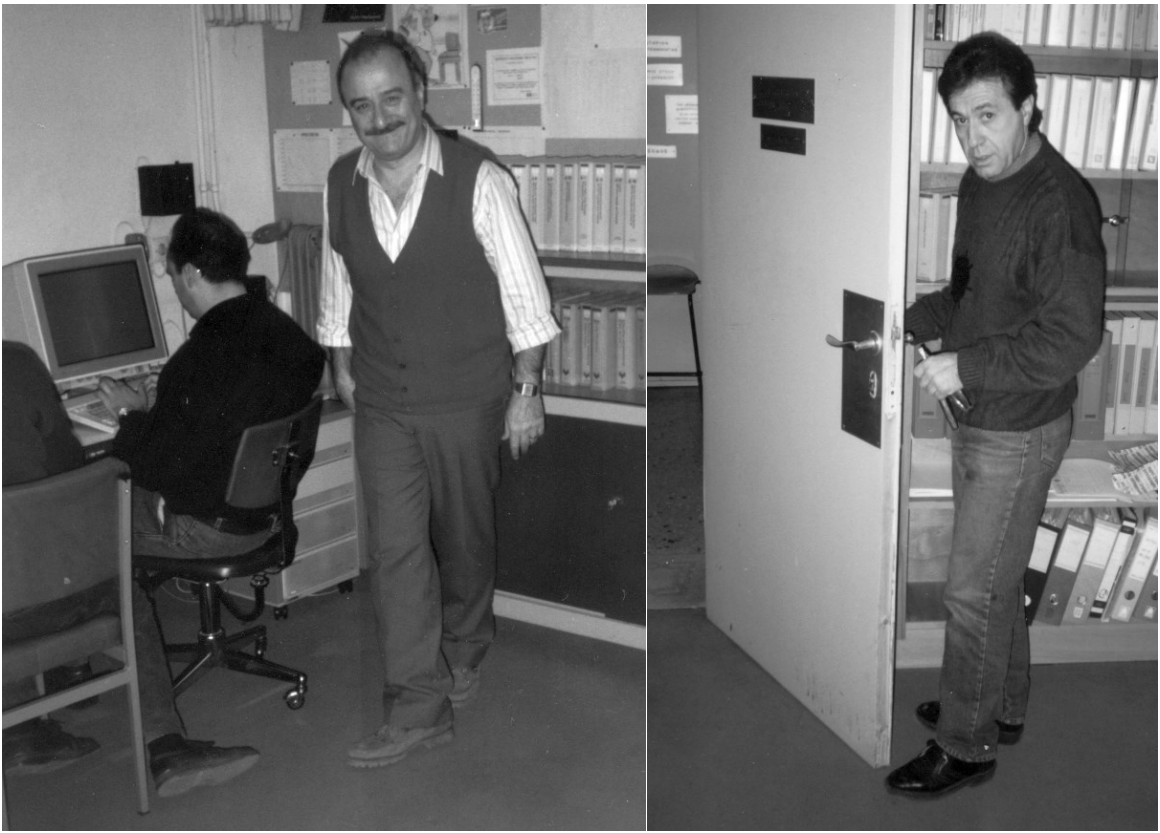

Εικόνα 19. Αριστερά διακρίνεται το μέλος ΕΕΠ του ΕΠΤ-ΕΜΠ Ν. Δημητρακόπουλος και δεξιά το μέλος ΕΔΤΠ-ΥΕ του ΕΠΤ-ΕΜΠ Β. Ηλίας. Και οι δύο υπηρετούσαν στο ΕΠΤ-ΕΜΠ την περίοδο του ατυχήματος στο Chernobyl. Στην αριστερή φωτογραφία διακρίνεται καθιστός ο εκλιπών Αναπληρωτής Καθηγητής του ΕΠΤ-ΕΜΠ Ε.Π. Χίνης, ως φοιτητής που εκπονούσε Διπλωματική Εργασία το 1986.

Υπάρχουν πολλά ακόμη πρόσωπα που ανήκουν στην ίδια ιστορία. Η επιλογή σχετικά με το ποια "πρέπει" να προστεθούν σε αυτήν την αφήγηση είναι ασφαλώς πολύ δύσκολη. Παρόλα αυτά δύο ακόμα είναι προβεβλημένα τμήματά της. Στην Εικόνα 20 διακρίνεται ο Δρ. Πανα-γιώτης Κρητίδης, επικεφαλής του Εργαστηρίου Ραδιενέργειας Περιβάλλοντος του Κέντρου Ερευνών "Δημόκριτος", στον οποίο έχει γίνει ήδη αναφορά. Στα αριστερά ο Σ.Ε. Σιμόπουλος. Θρυλικές παραμένουν οι, επιστημονικές ασφαλώς, διαφωνίες τους σχετικά με το πλήθος των δειγμάτων εδάφους που αρκούν για την αποτύπωση μιας καταστάσεως ραδιενεργού ρύπανσης καθώς και σχετικά με το ορθό βάθος δειγματοληψίας. Στην Εικόνα 21 διακρίνεται ο Δρ. Κωνσταντίνος Παπαστεφάνου, Καθηγητής Πυρηνικής Φυσικής στο ΑΠΘ, στον οποίο έχει γίνει ήδη αναφορά. Ο Δρ. Κ. Παπαστεφάνου είναι ένας ευγενώς αμιλλώμενος επιστήμονας και από τους πόλους της σχετικής έρευνας στη Βόρεια Ελλάδα.

## 6. Επίλογος

Για να μην εννοούνται ακουσίως δυσάρεστες επιπτώσεις στην Ελλάδα, από το ατύχημα στο Chernobyl, πρέπει στο σημείο αυτό να τονισθεί ότι σύμφωνα με τη μελέτη που έγινε στο ΕΠΤ-ΕΜΠ εύλογο διάστημα μετά τις δειγματοληψίες (Γεροντίδου, 1994) διαπιστώθηκε ότι: Κάτοικος που διαμένει στην περισσότερο ρυπασμένη περιοχή της χώρας για 50 συνεχόμενα έτη, εργάζεται για όλα αυτά τα έτη στο ρυπασμένο έδαφος ως αγρότης, κτηνοτρόφος ή τεχνικός, τρέφεται από τροφές που παράγονται αποκλειστικά στο ρυπασμένο έδαφος ή με ζώα που τράφηκαν από το ρυπασμένο έδαφος, αυτός λαμβάνει συνολική δόση εξαιτίας της ρύπανσης ίση με 10 mSv. Η αντίστοιχη δόση που λαμβάνει για το ίδιο διάστημα, ο ίδιος κάτοικος από το φυσικά ραδιενεργό περιβάλλον, το οποίο σε καμία περιοχή του πλανήτη, δεν μπορεί κανείς να αποφύγει, είναι περισσότερη από 200 mSv, δηλαδή 20 φορές μεγαλύτερη.

Είναι ευτύχημα ότι μετά το 1986, η χώρα δεν αντιμετώπισε ραδιενεργή ρύπανση από ατύχημα σε πυρηνικό αντιδραστήρα. Η αντίδρασή της, συνολικά, το 1986 έδειξε ότι υπήρξαν οι τεχνικές δυνατότητες και οι ικανοί επιστήμονες και τεχνικοί, οι οποίοι μπόρεσαν να αποτυπώσουν με καλή επάρκεια τη ρύπανση από το ατύχημα στο Chernobyl. Ασφαλώς πολλά έγιναν με προσωπική πρωτοβουλία και για άλλα υπήρξε πρόχειρη αντίδραση ή κακή οργάνωση. Αν συμβεί στο μέλλον κάτι παρόμοιο με το ατύχημα στο Chernobyl, πιστεύεται ότι η Ελλάδα είναι περισσότερο οργανωμένη, καλύτερα έτοιμη και επαρκώς στελεχωμένη για να την αντιμετωπίσει. Σε αυτό έχουν βοηθήσει και οι παρεμβάσεις της Ελληνικής Πολιτείας, οι οποίες έχουν ήδη απογραφεί. Χρειάζεται όμως να τονισθεί ότι σήμερα (2022) τα συνεργαζόμενα με την ΕΕΑΕ Ελληνικά Εργαστήρια έχουν περιορισθεί σημαντικά. Ενεργά παραμένουν, το ΕΠΤ-ΕΜΠ, το Εργαστήριο Πυρηνικής Φυσικής στο ΑΠΘ και το Εργαστήριο Πυρηνικής Φυσικής στο Πανεπιστήμιο Ιωαννίνων. Έλλειψη προσωπικού και αβέβαιο μέλλον έχουν το Εργαστήριο Πυρηνικής Τεχνολογίας του ΑΠΘ και το αντίστοιχο του Δημοκρίτειου Πανεπιστημίου Θράκης. Όλα συνολικά τα αναφερόμενα συνεργαζόμενα με την ΕΕΑΕ εργαστήρια, αντιμετωπίζουν συντριπτική για την ύπαρξή τους έλλειψη πιστώσεων. Ελπίζεται, ότι αυτές οι εμφανείς πλέον δυσκολίες να μην εμποδίσουν αυτές τις μονάδες να συμβάλουν με τον τρόπο που θα μπορούσαν αν παραστεί ανάγκη. Οδηγός είναι εξάλλου και τα όσα κατόρθωσε σχετικά το 1986 ο Σ.Ε. Σιμόπουλος με ελάχιστα μέσα και αμέτρητο αλτρουισμό. Σε ό,τι αφορά ειδικά στο ΕΠΤ-ΕΜΠ, το διάστημα από το 1986 έως το 2010, καταβλήθηκε από όλα τα μέλη του μια δύσκολο να απογραφεί προσπάθεια ώστε να αποκτηθεί κατάλληλος εξοπλισμός σχετικός με τέτοια ή παρόμοια έρευνα. Οι σχετικές διατάξεις θεωρούνται σήμερα ικανοποιητικά πλήρεις παρόλο που η τακτική χρηματοδότησή τους έχει πρακτικά σταματήσει λόγω των μέτρων λιτότητας. Είθε τέτοια έρευνα να μην απαιτηθεί ξανά.

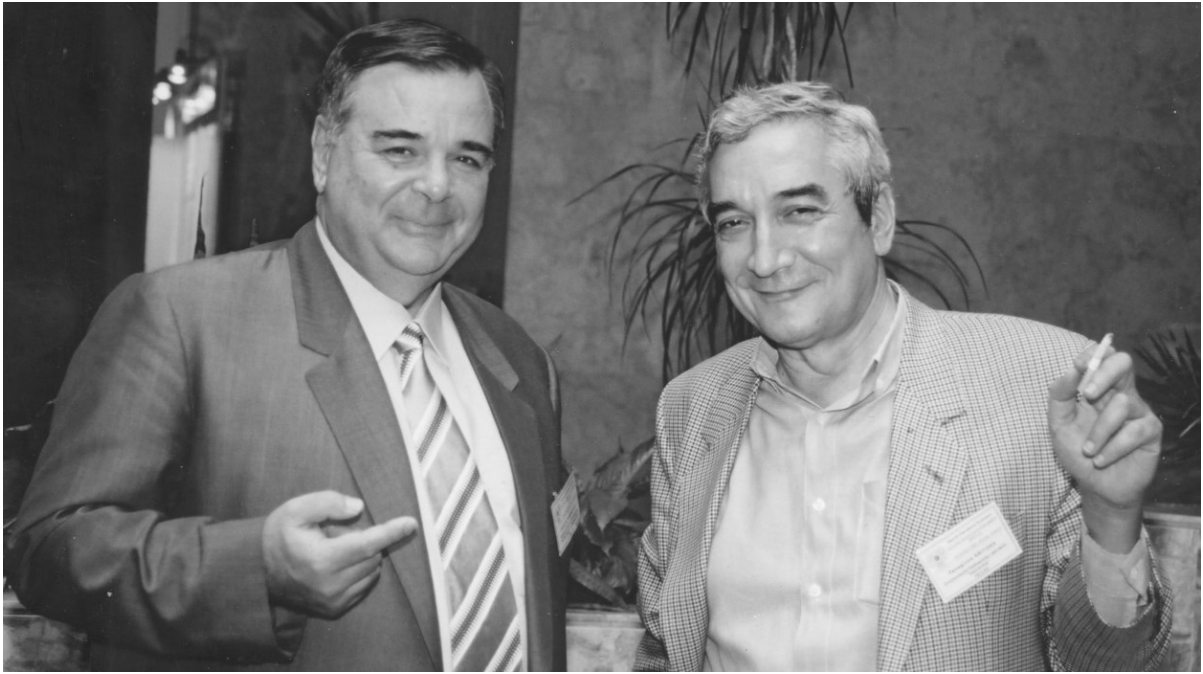

Εικόνα 19. Δεξιά ο, και μανιώδης καπνιστής, Δρ. Π. Κρητίδης παρέα με τον Σ.Ε. Σιμόπουλο σε ανέμελη χαλαρή στιγμή περί το 2002

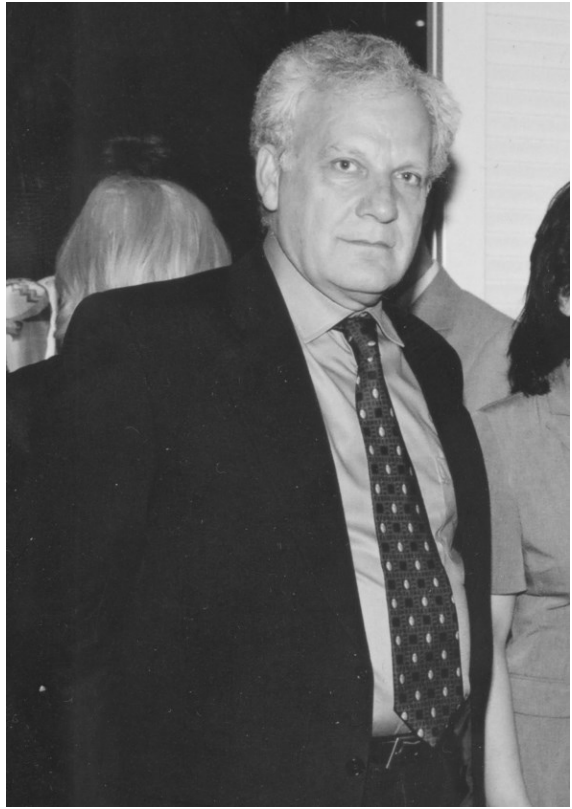

Εικόνα 20. Ο τ. Καθηγητής Πυρηνικής Φυσικής του ΑΠΘ Κ. Παπαστεφάνου

Βιβλιογραφία
Γεροντίδου Μ., *"Δοσιμετρική μελέτη του Ελληνικού πληθυσμού εξαιτίας του Cs-137 από το ατύχημα στο Chernobyl"*, Διπλωματική Εργασία, Τομέας Πυρηνικής Τεχνολογίας, Σχολή Μηχανολόγων Μηχανικών, Εθνικό Μετσόβιο Πολυτεχνείο, Αθήνα 1994

Ζαραφίδης Δ. και Παυλουδάκης Φ., *"Ο λιγνίτης της Πτολεμαΐδας, ιστορία, παρούσα κατάσταση και προοπτικές"*, ΔΕΗ: Λιγνιτικό κέντρο Δυτικής Μακεδονίας, 2008 (παρουσίαση που υπάρχει στο δικτυακό τόπο ptolemaida.gr)

Σιμόπουλος Σ.Ε., *"Αποτελέσματα προσδιορισμού ραδιενέργειας Cs-137, 1244 δειγμάτων χώματος εδάφους της Ηπειρωτικής Ελλάδας (Μάιος - Νοέμβριος 1986)"*, Έκθεση - Αναφορά ΜΡΧ-1, Τομέας Πυρηνικής Τεχνολογίας, Σχολή Μηχανολόγων Μηχανικών, Εθνικό Μετσόβιο Πολυτεχνείο, Αθήνα 1986 (τεκμήριο που διακινήθηκε εμπιστευτικά και βρίσκεται στο αρχείο του ΕΠΤ-ΕΜΠ)

Σιμόπουλος Σ.Ε., *"Μετρήσεις ραδιενέργειας των Ελληνικών εδαφών μετά το ατύχημα στο Chernobyl"*, Έκθεση - Αναφορά ΜΡΧ-2, Τομέας Πυρηνικής Τεχνολογίας, Σχολή Μηχανολόγων Μηχανικών, Εθνικό Μετσόβιο Πολυτεχνείο, Αθήνα 1987 (τεκμήριο που διακινήθηκε εμπιστευτικά και βρίσκεται στο αρχείο του ΕΠΤ-ΕΜΠ)

De Cort M. et al., *"Atlas of Caesium deposition on Europe after the Chernobyl accident"*, Luxembourg, Office for Official Publications of the European Communities, ISBN 92-828-3140-X, 1998

DEMO 86/3 G, *"The Chernobyl Nuclear Accident and its Consequences for Greece"*, Report No I. Greek Atomic Energy Commission, Nuclear Research Centre Democritos, Athens, Greece, 1986.

Petropoulos N.P., Anagnostakis M.J., Hinis E.P., Simopoulos S.E., *"Geographical mapping and associated fractal analysis of the long-lived Chernobyl fallout radionuclides in Greece"*, Journal of Environmental Radioactivity, 53(1): 59-66, 2001

Simopoulos S.E., *"Soil sampling and Cs-137 analysis of the Chernobyl fallout in Greece"*, International Journal of Radiation Applications and Instrumentation. Part A. Applied Radiation and Isotopes, 40(7): 607-613, 1989

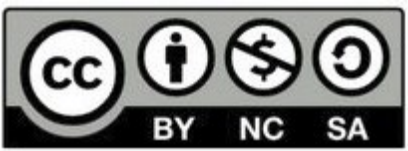